\documentstyle[12pt]{article}
\def\np#1,#2,#3 {Nucl.\ Phys.\ {\bf B#1} \up(19#2\up) #3}
\def\pl#1,#2,#3 {Phys.\ Lett.\ {\bf #1B} \up(19#2\up) #3}
\def\sovj#1,#2,#3 {Sov.\ J.\ Nucl.\ Phys.\ {\bf #1} \up(19#2\up) #3}
\def\jetp#1,#2,#3 {Sov.\ Phys.\ JETP {\bf #1} \up(19#2\up) #3}
\def\rmp#1,#2,#3 {Rev.\ Mod.\ Phys.\ {\bf #1} \up(19#2\up) #3}
\def\prd#1,#2,#3 {Phys.\ Rev.\ {\bf D#1} \up(19#2\up) #3}
\def\prl#1,#2,#3 {Phys.\ Rev.\ Lett.\ {\bf #1} \up(19#2\up) #3}
\def\physrep#1,#2,#3 {Phys.\ Rep.\ {\bf #1} \up(19#2\up) #3}
\def\journal#1,#2,#3,#4 {#1 {\bf #2} \up(19#3\up) #4}
\def\preprint#1,#2,#3 {#1 preprint #2 \up(19#3\up)}

\begin{document}
\begin{titlepage}
\vskip 15 mm
\title{Neutrino properties from maximally-predictive GUT models \\
and the structure of the heavy Majorana sector.}
\vskip 10 mm
\author{\bf E. Papageorgiu}
\vskip 2 mm
\centerline{Laboratoire de Physique Th\'{e}orique et Hautes Energies}
\centerline{Universit\'{e} de Paris XI, B\^{a}timent 211, 911405 Orsay, France}
\vskip 20 mm
\abstract{Starting from a complete set of possible parametrisations of the
quark-mass matrices that have the maximum number of texture
zeros at the grand unification scale, and the Georgi-Jarlskog mass relations,
we classify the neutrino spectra with respect to the unknown structure of the
heavy Majorana sector.
The results can be casted into a small number of phenomenologically
distinct classes of neutrino spectra, characterised by universal mass-hierarchy
and oscillation patterns.
One finds that the neutrino masses reflect the natural hierarchy among
the three generations and obey the quadratic seesaw, for most GUT models
that contain a rather unsophisticated Majorana sector. A scenario with
$\nu_{\tau}$ as the missing hot dark matter component and $\nu_e\leftrightarrow
\nu_{\mu}$ oscillations accounting for the solar neutrino deficit comes
naturally out of this type of models and is
very close to the experimental limit of confirmation or exclusion.
In contrast, in the presence of a strong hierarchy of heavy scales
or/and some extra symmetries in the Majorana mass matrix, this natural
hierarchy gets distorted or even reversed. This fact can become a link
between searches for neutrino oscillations and searches for discrete symmetries
close to the Planck scale.}
\vskip 1 truecm
\noindent {\bf LPTHE Orsay 45}
\end{titlepage}

\section{Introduction.}
The quest for understanding the Yukawa sector of the Standard Model (SM), which
could mean
finding, as a first step, simple fermion-mass and quark-mixing relations among
the members of the three known families, represents an equally significant
challenge to the standard model (SM) as the prospect of unification of the
three gauge
couplings within the scope of a more fundamental theory.
As a matter of fact, the two problems are related to each other, as most
grand-unified (GUT) models
imply also some partial unification of Yukawa couplings. The by now famous
$m_b = m_{\tau}$ equality between the mass of the bottom quark and the
corresponding charged lepton of the third family at $M_{G}$, the scale of grand
unification,
has been one of the early successes of minimal SU(5) \cite{GG}.
For the first two families, where it is empirically known that
$m_d/m_e\,\simeq\, 10\, m_s/m_{\mu}$, rather than $\simeq m_s/m_{\mu}$,
the hope was (and still is) that some other effect may modify these simple mass
relations, without
significantly altering the two-Yukawa coupling unification scheme.
A most promissing attempt in this direction has been the {\it Ansatz} of Georgi
and
Jarlskog (GJ)\cite{GJ}, subsequently implemented also into other (so-called
predictive) GUT models [3]-[8].
In some models based on the $SO(10)$ group one is even led towards
a unification of all three Yukawa couplings of e.g. the third
family: \cite{Lang} $h_t(M_G) = h_b(M_G) = h_{\tau}(M_G)$.

The structure of the three-generation
Yukawa matrices, which
parametrise the couplings of the fermions to the Higgs sector, is
commonly attributed to the
existence of $U(1)$ - axial horizontal symmetries, broken at some
intermediate scale between the electroweak and the unification scale,
or/and to the existence of discrete symmetries that often appear after
compactification of the superstring \cite{discrete}.
In the hope of finding some fundamental symmetry of this type,
 a different approach has been put forward lately \cite{{DHR},{Giudice},{RRR}}.
Instead of
trying to accomodate the empirically known mass and mixing parameters
in different GUT models, the procedure has been to first find appropriate
{\it Ans\"atze} for the structure of the Yukawa matrices at $M_{G}$ which give
the
correct values at low energies. In order to limit the number of possible
choices, and be predictive with respect to an expected improvement of the
experimentalBB values of the 13 mass- and mixing-parameters of the SM in the
near future, the principle of economicity has again
been applied, meaning as few input parameters as possible.
One way to meet this requirement is namely to have as many zero entries
as possible
and/or some extra symmetry, e.g. in flavour space, as this is natural
for models of the $SO(10)$ group.
{\it Ans\"atze} of this type have been made in the past by Fritzsch
\cite{Fritzsch} and
Stech \cite{Stech} for the quark mass matrices at the electroweak scale.
More recently, Dimopoulos, Hall and Raby (DHR)\cite{DHR} and Giudice
\cite{Giudice}
have proposed new parametrisations of the Yukawa sector at $M_G$  with
seven or
six parameters only, to describe the six quark masses, the three mixing
angles
and the CP-violating phase of the Cabibbo-Kobayashi-Maskawa (CKM) matrix.
For the running of the Yukawa couplings between $M_G$ and the low energy scale,
additional assumptions on the Higgs sector and particle content of the
theory
are needed. Since it was shown that precision data from LEP are consistent with
unification of the three gauge couplings within the minimal supersymmetric
standard model
(MSSM) \cite{RR}, it has become the most popular candidate for a description of
the
physics between the electroweak scale $M_Z$ and the unification scale $M_G$,
which in this context turns out to be of the order of $M_G\simeq 10^{16}$ GeV.
However, as most SUSY-GUT models
share the same particle content with the MSSM at energies below $M_{G}$, while
all nonsupersymmetric GUTs suffer from the so-called hierarchy problem, this
choice is representative for the evolution of a whole class of models
\cite{Lang}.

In a recent paper Ramond, Roberts and Ross (RRR)\cite{RRR} have
reversed this approach and, starting from what is measured at low energy,
have
provided a classification of all possible sets of
quark-Yukawa matrices, which are hermitian, i.e. symmetric in flavour space,
and have
five or six texture zeros at $M_G$.
In this way a unified picture in terms of a perturbative generation of
the quark-Yukawa sector at the scale $M_G$ has been achieved, which
incorporates the Fritzsch, the DHR, and the Giudice {\it Ans\"atze}, and sets
the level
of accuracy needed to discriminate between them by improved measurements of
the CKM matrix elements.

By means of the Georgi-Jarlskog Ansatz, which successfully relates the
Yukawa couplings of
the charged leptons to those of the down quarks at $M_{G}$,
the DHR \cite{DHRN} and other groups \cite{Leonta} have in addition to the
quark sector been able to make predictions
concerning also the existence of mixing in the lepton sector, if the three
ordinary left-handed neutrinos of the SM
were to obtain a small mass through mixing with extra heavy (right-handed)
neutrino-like states by the so-called seesaw mechanism. Predictions for this
sector are possible only
because the very idea of grand unification naturally implies some
proportionality relations among the
Higgs-Yukawa couplings of the fermions belonging to the same multiplet. Thus
in models with an $SO(10)$ symmetry, the neutrino-Yukawa couplings which are of
the Dirac type are usually proportional to those for the up-type quarks.
However the structure of the Majorana-type Yukawa matrix, responsible for
giving large masses to the right-handed neutrino states, thus leading to
seesaw-suppressed values for the masses of the ordinary neutrinos,
is in general not known
\footnote{For counter examples see refs.(\cite{DHRN}, \cite{Elena},
\cite{Babu}).}
and has been chosen, as a matter of convenience, to be diagonal or
proportional to the up-quark (or down-quark) mass matrices \cite{Leonta},
\cite{Ellis}, \cite{Langacker}. Due to its fundamental importance in providing
the only known mechanism for suppressing the unacceptably large neutrino masses
implied by most GUT models, one would like
to know the full phenomenological impact also of this Yukawa sector, whose
scale and structure are most likely determined by the physics at the Planck
scale. Moreover, it seems to be the case that different models give similar
neutrino spectra \cite{DHRN}, \cite{Leonta}, pointing towards some universality
that may follow on one hand from the simple GUT relations and on the other hand
from some extra symmetries at the Planck scale. It could well be that a new
classification
scheme with respect to the structure of the heavy Majorana-mass sector and its
predictions for the masses and the mixing of the leptons could shed some more
light on the latter.
With the prospect of getting close to probing interesting regions of $\Delta
m^2$, the mass-difference
squared of two neutrino species, and $sin^2 2\theta$, the parameter which gives
their mutual mixing, in coming neutrino-oscillation experiments
(the NOMAD, the CHORUS, the ICARUS and other proposals) \cite{Rubia}, the
possibility of testing one of the basic ideas of grand unification through a
classification of such maximally predictive GUT models could become an
interesting project. This paper will be devoted to these
questions.

We will start in section (2) with a review of the expectations
on neutrino masses and mixing from theory, experiment
and observation and discuss the common prejudices.
We will then in section (3) extend the RRR approach to the lepton sector
such that the structure of the charged-lepton Yukawa matrix is fixed by the GJ
relations to the matrix of the down-type quarks and the one of the
Dirac-neutrino states to the up-type quarks as implied by the simplest
implementation of the GUT idea. The unknown heavy Majorana-neutrino
sector is first chosen to be arbitrary and general, to be subsequently
classified
according to the mass- and mixing-patterns it leads to after diagonalisation
of the effective neutrino mass matrix.
In section (4) we will examine four cases leading to universal neutrino mass
and mixing patterns characterised by a specific hierarchical order. In section
(5), we discuss the implications for the coming neutrino oscillation
experiments, and in section(6), we give the conclusions.

\section{Massive neutrinos between hope and prejudice.}

The absence of a gauge or any other principle that could justify a zero mass
for the neutrino as it does for the photon may be one of the strongest
theoretical prejudices in favour of massive neutrinos. The exasperation with
the ``overly'' successful Standard Model and a general attitude of
looking forward to seeing ``new physics'' has been the prejudice's driving
force. Three neutrino-deficit phenomena based on astrophysical observation and
cosmological considerations have become the nurishing substance of this
hope. The three experimental uper bounds on the masses of the tau- muon- and
electron neutrino of $32$ MeV, $270$ keV and $1$ eV respectively have along
with other data on neutrino oscillations and double-beta decay severly limited
the range of beyond-the-SM speculations \cite{Spiro}.
The latter have also helped to articulate new questions, as to why for example
the electron neutrino is so much lighter than the electron -at least some five
orders of magnitude- , adding some extra power to the mass-hierarchy puzzle, to
which the seesaw mechanism \cite{seesaw}
became a common reply.

\subsection{The seesaw mechanism.}
The idea of the seesaw, first implemented in partially or completely unified
theories with a left-right symmetry such as $SO(10)$ \cite{seesaw},
is based upon the simple fact that for a mass matrix of the type:
\begin{equation}
M = \left(
\begin{array}{cc}
0 & a \\
a & b \\
\end{array}
\right) \,,
\end{equation}
where $a \ll b$, a simple rotation leads to eigenvalues with a large mass
splitting:
\begin{equation}\label{ab}
m_1 \sim {a^2\over b}  \qquad    m_2 \sim b \,,
\end{equation}
and a simple mass and mixing-angle relation:
\begin{equation}
{1\over 2}\, tan 2\theta \simeq {a\over b} \,,
\end{equation}
and therefore to:
\begin{equation}\label{theta}
sin \theta \sim \theta \sim \sqrt{{m_1\over m_2}} \,.
\end{equation}
So the main lesson of this trivial exercise is the twofold quadratic scaling
behaviour of the mass ratio $m_1/m_2$, that is at the same time proportional to
$a^2/b^2$ and to $sin^2 \theta $.

When $M$ represents the mass matrix for one generation of neutrinos, written in
the left-  and right-handed neutrino basis $(\nu, N^c)$, $a$ is a small Dirac
mass $m_D$ that is proportional to a quark or lepton mass, and $b$ a large
Majorana mass $R$ for the right-handed neutrino that is proportional to some
scale $M_X \gg M_Z$ :
\begin{equation}
M_{\nu} = \left(
\begin{array}{cc}
0 & m_D \\
m_D & R \\
\end{array}
\right) \,.
\end{equation}
The fact that the entry in the upper left corner of $M_{\nu}$ is usually zero,
signifies the absence of an $SU(2)_L$ Higgs triplet that could  give a Majorana
mass also to the left-handed neutrino. It is interesting to note that the
``seesaw'' notion, {\it i.e.} that the mixing of a light state with a heavy
state in the mass matrix can render the former even lighter, is employed only
in the  context of a Majorana-neutrino sector and/or in the presence of heavy
singlets, mixing with the quarks \cite{USS}, but never in the context of the
quark-Yukawa matrices which are nevertheless of the same perturbative type (see
{\it e.g} equ.(33)

In the case of a two-generation or three-generation neutrino mixing, where
$m_D$ and $R$ are replaced by the corresponding $2\times 2$ or $3\times 3$
matrices $M_u$ and $M_R$, one finds similar relations among the masses and the
mixing of the light neutrinos if, after
block diagonalisation of:
\begin{equation}\label{Mnu}
M_{\nu} = \left(
\begin{array}{cc}
0 & M_u \\
M_u^T & M_R \\
\end{array}
\right) \,,
\end{equation}
and assuming that the matrix $M_R$ is not singular, the effective
light-neutrino mass matrix:
\begin{equation}
M_{\nu}^{eff} \simeq M_u \, M_R^{-1} \, M_u^T \,,
\end{equation}
is fully or partly of the same perturbative type as $M$
\footnote{When $M_u$ is chosen to be a real matrix, as this is a common choice
for the mass matrix of the up-quark sector, $M_u^T$ is its transpose.
In general though, $M_{\nu}$ might not even be a hermitian matrix, in which
case one will have to diagonalise the matrix $M_{\nu} M_{\nu}^{\dag}$ which is
hermitian for any complex matrix $M_{\nu}$.}.
In this case however, the light-neutrino  mass eigenvalues will in general be
proportional to a more complicated function of the entries of the matrices
$M_u$ and $M_R$. Considering the case of  weak intergenerational mixings in
$M_R$
and a perturbative structure ($\grave{a}\, la$ Fritzsch) for $M_u$, the ratio
of any two light neutrino masses exhibits a simple scaling behaviour:
\begin{equation}\label{SS}
{m_{{\nu}(x)}\over m_{{\nu}(y)}} = {(m_{u_i} m_{u_j})_{(x)}\over (m_{u_k}
m_{u_l})_{(y)}} \cdot {R_{m (y)}\over R_{n (x)}} \cdot a_{(xy)} \,,
\end{equation}
where $i ... n$ are generation indices, $x$ and $y$ label the two neutrino
species, and $a_{(xy)}$ may be any ratio of additional heavy Majorana scales
participating in the seesaw.

Now in the case where $M_R$ has only one scale, equ.(\ref{SS}) reduces either
to the quadratic seesaw:
\begin{equation}\label{QSS}
{m_{\nu (x)}\over m_{\nu (y)}} = {m_{u_i}^2 \over m_{u_j}^2} \,,
\end{equation}
or to the linear seesaw:
\begin{equation}\label{LSS}
{m_{\nu (x)}\over m_{\nu (y)}} = {m_{u_i} \over m_{u_j}} \,,
 \end{equation}
or to the mixed case:
\begin{equation}\label{MSS}
{m_{\nu (x)}\over m_{\nu (y)}} = {m_{u_i} m_{u_j} \over m_{u_k}^2} \,.
\end{equation}
Futhermore it is easy to check that when the matrix $M_R$ can be diagonalised
simultaneously with the matrix $M_u$ one obtains the quadratic seesaw
with a neutrino mass hierarchy corresponding to the quark or charged-lepton
mass hierarchy, {\it i.e.} $x = i$ and $y = j$, and a simple proportionality
relation between the quark- and lepton-mixing matrices:
\begin{equation}
V_l \simeq V_{CKM} \,.
\end{equation}
This is indeed the simplest version of seesaw that one can have.
In contrast, in the presence of a strong hierarchy and/or strong
intergeneration mixings in $M_R$, one would rather expect a distortion of this
rather ``natural'' hierarchy pattern, up to the point that it actually gets
reversed. This is an interesting possibility to explore that would have
important implications for future neutrino-oscillation experiments
\cite{Petcov}.

\subsection{The neutrino-deficit problem.}

Let us now come to the observed neutrino deficits. The most prominent one,
since it has been confirmed by four different experiments involving three
different targets ($Cl^{37}$, $e^{-}$ in $H_20$, and $Ga^{71}$) \cite{Ga},
is the solar neutrino deficit:
The flux of the neutrinos coming from the sun and measured sofar is half
to one third of the expected number of SNUs calculated by two groups using the
standard solar model (SSM) \cite{SSM}.
Apart from a ever decreasing probability of resolving this discrepancy by
altering the parameters of the SSM [26] and while waiting for the calibration
of the two $Ga^{71}$ experiments, the most promissing solution seems to be
the Pontecorvo or the Mikheyev-Smirnov-Wolfenstein (MSW) mechanism of
vacuum or matter-enhanced oscillations of the electron neutrinos from the sun
into some other neutrino species in their way to the earth \cite{Ponte}.
Analysing the latest data of the four experiments Krastev and Petcov have found
three disconnected areas in the $\Delta m^2$ and $sin^2 2\theta$ plot
where such two- or even three-neutrino-flavour oscillations are allowed
\cite{Krastev}.
The vacuum-oscillation solution (VOS) is characterized by very small mass
differencies and a large mixing angle:
\begin{equation}\label{VOS}
\begin{array}{c}
\Delta m^2 \simeq (0.5 - 1.) \times 10^{-10} eV^2 \\
sin^2 2\theta_{\nu_e - \nu_x} \simeq 0.75 - 1.\,,
\end{array}
\end{equation}
while in the case of matter-enhanced oscillations there is a small-angle
non-adiabatic MSW solution:
\begin{equation}\label{MSW}
\begin{array}{c}
\Delta m^2 \simeq (0.3 - 1.2)\times 10^{-5} eV^2 \\
sin^2 2\theta_{\nu_e - \nu_x}  \simeq (0.5 - 1.6) \times 10^{-2} \,,
\end{array}
\end{equation}
and a large angle solution:
\begin{equation}\label{MSW1}
\begin{array}{c}
\Delta m^2 \simeq (0.3 - 3.) \times 10^{-5} eV^2 \\
sin^2 2\theta_{\nu_e - \nu_x} \simeq 0.6 - 0.7 \,,
\end{array}
\end{equation}
where the upper range of $sin^2 2\theta$ in the last equation has been reduced
from $0.9$ to $\simeq 0.7$ due to the non-observed effect of double conversion
of the electron neutrinos from the supernova SN87A \cite{Smirnov}.
The quoted numbers strictly hold for transitions of $\nu_e$'s into
$\nu_{\mu}$'s or $\nu_{\tau}$'s. For a transition into a sterile neutrino state
(with respect to the electroweak interaction) $\nu_s$, the allowed range for
$sin^2 2\theta$ shrinks to smaller values in the case of a small-angle solution
and to larger values in the case of a large-angle MSW solution \cite{Krastev}.

Compared to the Cabibbo quark-mixing angle $sin \theta_c \equiv s_{12} \simeq
0.22$, the value of the neutrino-mixing suggested by the small-angle MSW
solution
is smaller and could indeed correspond to a lepton-mixing angle:
\footnote{We use the following short-hand notations: $sin \theta_{ij} \equiv
s_{ij}$ and $cos \theta_{ij} \equiv c_{ij}$.}
$s_{12}^l \simeq 0.04 - 0.07$ $\simeq\sqrt{{m_e\over m_{\mu}}}$,
unless it represents the mixing between the first and third generation, in
which case it is larger than the corresponding one in the quark sector. In
contrast, the values suggested by the large-angle solution always
exceed the values of the CKM matrix elements and, given the very small mass
differences they correspond to, they would rather represent an
anomaly in any attempt to find universal relations for all fermion sectors
among the fermion masses and the mixing angles.
A second general remark concerns the range of the masses that one would expect
for neutrinos participating in such oscillations.
Unless there is a symmetry to guarrantee small mass differences between masses
whose scale lies much higher, and in order to avoid fine tuning, one would
expect at least one of the neutrinos to have a mass in the range of
$m_{\nu_i} \simeq (\Delta m^2)^{1/2}$, which in the case of matter-enhanced
oscillations implies a mass scale $\Lambda_1\simeq (2 - 3)\times10^{-3}$ eV.

A second neutrino deficit - a more controversial topic due to the large
experimental error bars and counter evidence from two experiments -  has been
reported by several experimental groups \cite{IMB} with respect to the expected
ratio of muon- to electron-neutrino flux produced by hadronic collisions in the
upper atmosphere and measured deep underground. Again a straight forward
explanation of this
can be given if one assumes that the muon neutrinos oscillate into other
light-neutrino species with the values of:
\begin{equation}\label{AN}
\begin{array}{c}
\Delta m^2 \simeq (0.5 - 0.005) eV^2 \\
sin^2 2\theta_{\nu_{\mu} - \nu_x} \simeq 0.5 \qquad .
\end{array}
\end{equation}
Since the option of a $\nu_{\mu} - \nu_e$ oscillation in the range of
$\Delta m^2 < 0.007 eV^2$ for $sin^2 2\theta \simeq 1$ and of large
$\Delta m^2$ for $sin^2 2\theta \leq 4 \times 10^{-3}$ has been excluded
from reactor and accelerator experiments \cite{Acclim}, there remain only two
possibilities, {\it i.e.} a transition into a tau neutrino or a sterile
neutrino-like state. However the possibility of $\nu_{\mu}$ or any other active
neutrino component
oscillating into a light sterile neutrino state according to the range of
parameters suggested by equ.(\ref{AN})
seems to be excluded from the data on nucleosynthesis, which imply that the
effective neutrino degrees of freedom consistent with the measured $H_e$
abundance is less than $3.3$, and lead to a constraint of \cite{Mark}:
$$\Delta m^2 \, sin^2 2\theta \leq 3.6 \times 10^{-6} eV^2 \,.$$
It should also be noted that the latest results of the BAKSAN and IMB
background measurements of upcoming muons seem to narrow down the allowed
range for such oscillations to a vanishingly small area \cite{Spiro}.

Now if both neutrino deficits were to be confirmed and interpreted as neutrino
oscillations between massive SM neutrinos this would naturally imply the
following mass hierarchy in the neutrino sector: $ m_{\nu_e} \ll m_{\nu_{\mu}}
\sim 10^{-3}$ eV $< m_{\nu_{\tau}}\sim 10^{-1}$ eV.
This scenario is however at odds with the possibility of resolving as well the
third observed deficit, known as the hot dark matter (HDM) problem,
in a scenario with three light neutrinos only. The COBE results on the
anisotropy of the cosmic microwave background radiation and the data on the
angular correlations of galaxies and galactic clusters can be best fitted by a
``coctail'' of $70 \%$ cold dark matter and
$30 \%$ hot dark matter \cite{Silk}, the latter consisting of neutrinos with
mass of a few electronvolts:
\begin{equation}\label{HDM}
\Sigma_i m_{\nu i} \simeq 7 eV \,,
\end{equation}
the sum being over all light neutrino components.
Assuming a ``natural-hierarchy'' scenario for the neutrino masses
\cite{Petcov}, to be discussed in more detail later on, this would point
towards a $7$ eV tau neutrino which can therefore not fulfill its role as a
participant in a $\nu_{\mu} - \nu_{\tau}$ oscillation in the earth's atmosphere
\cite{Caldwell}.
If all three observed deficits were to be of the same origin, {\it i.e.}
related to non-standard neutrino properties and in particular to oscillations
between different mass eigenstates, another light (sterile) neutrino state
would be needed in order to avoid a fine-tuned mass matrix and one would be
left with two scenarios \cite{Caldwell}:
Either the three known neutrinos participate in oscillations which allow for a
resolution of the solar and atmospheric neutrino deficits so that a massive
sterile neutrino would be needed to account for the HDM, or the muon and
tau neutrinos have masses of a few electronvolts so as to be the main
components of the HDM and participate in oscillations such as required for the
atmospheric neutrino puzzle to be resolved, in which case there should exist
$\nu_e - \nu_s$
oscillations, responsible for the solar neutrino deficit.
It seems that in the first scenario it is very hard to reconcile the
nucleosynthesis constraint with the HDM requirement. On the other hand
the second scenario requires a slight fine tuning of the muon and tau
neutrino masses. So there is no scenario which would naturally satisfy all
three requirements.
Of course once the requirement of resolving all three puzzles via neutrino
masses and oscillations, which is by no means compelling, is dropped, more
scenarios are available to speculation.


Since the existence of an atmospheric neutrino deficit is experimentally
disputable we would like in what follows to focus on scenarios which could do
away only with the solar neutrino puzzle and provide the relativistic component
needed in order to resolve the dark matter problem, assuming the existence of
three light neutrinos only.
Let us denote by $m_{\nu_1}$, $m_{\nu_2}$ and $m_{\nu_3}$ the corresponding
mass eigenstates in increasing mass order. We first notice that in order to
satisfy both requirements there should be at least two mass scales, {\it e.g.},
\footnote{For the moment we consider only the MSW solutions to the solar
neutrino problem.}
$\Lambda_1 \simeq (2 - 3) \times 10^{-3}$ eV
and $\Lambda_2 \simeq 7$ eV and thus a hierarchy of:
\begin{equation}
\Lambda\equiv {\Lambda_1 \over \Lambda_2} \sim (3 - 4) \times 10^{-4} \,.
\end{equation}
This hierarchy could then be realised in two different ways: One possibility
would be that the light-neutrino masses follow a similar hierarchy pattern as
the three up- or down-type quarks,
\begin{equation}\label{NH}
m_{\nu_e} \leq m_{\nu_{\mu}} \sim \Lambda_1 \ll m_{\nu_{\tau}}\sim \Lambda_2
\,,
\end{equation}
where $m_{\nu_{e,\mu}}$ is the dominant component of $m_{\nu_{1,2}}$ and
$m_{\nu_{\tau}}\simeq m_{\nu_3}$. We shall refer to this as the ``natural
hierarchy''pattern. The alternative then would be to have a
reversal of the natural hierarchy pattern, {\it i.e} to have:
\begin{equation}\label{RH1}
m_{\nu_e} \leq m_{\nu_{\tau}}\sim \Lambda_1  \ll m_{\nu_{\mu}}\sim \Lambda_2
\,,
\end{equation}
or
\footnote{Notice that the upper bound on the electron-neutrino mass is
such that it cannot at the same time be the heaviest neutrino and an HDM
candidate.}
\begin{equation}\label{RH2}
m_{\nu_{\tau}} \leq m_{\nu_e} \sim \Lambda_1  \ll m_{\nu_{\mu}}
\sim \Lambda_2 \,.
\end{equation}
Such an ``inverse hierarchy'' would naturally imply a strong $\nu_e -
\nu_{\tau}$ mixing.

We would like now to consider the possibility that there is indeed a natural
hierarchy among the neutrino masses and see whether it could more
likely originate from a quadratic or a linear seesaw mechanism.
Assuming that the heavy Majorana sector has only one scale $R$, the quadratic
seesaw would yield a neutrino-mass ratio in terms of the masses of the up-type
quarks,
\begin{equation}\label{QSS1}
{m_{\nu_2} \over m_{\nu_3}} \sim {m_c^2 \over m_t^2} \sim (0.5 - 3) \times
10^{-4} \simeq {\cal O}(\Lambda) \,,
\end{equation}
that is compatible with the value of $\Lambda$ as required by the two MSW
solutions to the solar neutrino problem, while
the linear seesaw in terms of masses of any of the three fermion sectors
leads to values that are considerably larger ($\sim 10^{-2}$), and the
quadratic seesaw in terms of down-quark or charged-lepton masses leads to
smaller values of order $10^{-3}$.
Despite the uncertainty coming from
the charm- and top-quark masses - evaluated at the electroweak scale $M_Z$ -,
clearly the hierarchy suggested by equ.(\ref{NH}) seems to follow from the
quadratic rather than the linear seesaw and involves the masses of the up
quarks.
On the other hand, if one would try to understand the small-angle MSW mixing in
equ.(\ref{MSW}) in terms of some power of the mass ratio of the corresponding
up-type quarks one would find that:
\begin{equation}
s_{12}^l \simeq \sqrt{{m_{\nu_1}\over m_{\nu_2}}}
\simeq \sqrt{{m_u \over m_c}}  \,,
\end{equation}
a relation which is typical for a linear seesaw.
Apparently the way neutrino masses are generated,  seems to differ when going
from the third to the second generation or from the second to the first.
The simplest mass pattern giving rise to such a scenario could {\it e.g.}
be:
\begin{eqnarray}\label{NS}
m_{\nu_1} & \sim & {m_u m_c \over R} \\
m_{\nu_2} & \sim & {m_c^2 \over R} \\
m_{\nu_3} & \sim & {m_t^2 \over R} \,.
\end{eqnarray}
Using again the naive mass and mixing-angle relations of equ.(\ref{theta}),
one can estimate
the remaining two neutrino-mixing angles and compare them to the corresponding
quark-mixing angles \cite{Danilov}:
\begin{eqnarray}\label{NA}
s_{23}^l \sim {m_c\over m_t} & < & s_{23} \sim 0.03 - 0.05 \\
s_{13}^l \sim \sqrt{{m_u m_c \over m_t^2}} & \ll & s_{13} \sim (2. - 7.) \times
10^{-3} \,.
\end{eqnarray}
The scale of the Majorana sector can be fixed by requiring that the heaviest
neutrino be the HDM candidate:
\begin{equation}\label{mt}
m_{\nu_3} \simeq m_{\nu_{\tau}} \simeq 7 eV \,.
\end{equation}
This gives:
\begin{equation}\label{R}
R \sim {\cal O}(10^{12})\, GeV \,,
\end{equation}
and masses to the other neutrinos:
\begin{eqnarray}\label{NM}
m_{\nu_1} & \sim & 10^{-5} eV \\
m_{\nu_2} & \sim & 10^{-3} eV \,.
\end{eqnarray}

We turn next to alternative scenarios, that could also provide a solution to
the solar neutrino problem and the HDM, but do not follow directly from
the simple seesaw relations, eqs.(\ref{QSS},\ref{LSS},\ref{MSS}).
Under this category fall the two large-angle solutions to the solar neutrino
problem, eqs.(\ref{VOS},\ref{MSW1}), simply because they indicate that the two
neutrinos participating in such oscillations are linear combinations of
mass-degenerate eigenstates:
$ \nu_{1/2} \simeq {1\over 2} (\nu_e \pm \nu_x) $
with $m_{\nu_1} \simeq m_{\nu_2}$. This is only possible in the context of a
more elaborate heavy Majorana sector, where for example $M_R$ or some of its
subdeterminants could be singular, and/or the presence of a large-scale
hierarchy can compensate the hierarchy in $M_u$.
Obviously the same is true also of any inverse-hierarchy scenario.
Therefore these options require a full treatment of the neutrino mass matrix
within the context of a particular model or at least of an {\it Ansatz}.

On the other hand, the semi-quantitative treatment of the natural-hierarchy
scenario should be also viewed with a great deal of caution,
for two main reasons: First because even in the quark sector the simple
relations we have employed are only valid in the approximation of
two-generation mixing. As an example we take the original Fritzsch model
\cite{Fritzsch}, where the mass matrices for the up- and down-type quarks were
parametrised according to:
\begin{equation}
M_F = \left(
\begin{array}{ccc}
0 & A & 0 \\
A & 0 & B \\
0 & B & C
\end{array}
\right) \,,
\end{equation}
with $A$, $B$, $C$ written in terms of the up-quark (down-quark) masses:
\begin{equation}
A \simeq \sqrt{m_{u(d)}\cdot m_{c(s)}} \ll
B \simeq \sqrt{m_{c(s)}\cdot m_{t(b)}} \ll
C \simeq m_{t(b)} \,.
\end{equation}
For the large Cabibbo-mixing angle one does indeed recover the simple
Gatto-Sartori-Tonin-Oakes relation \cite{GSTO}:
\begin{equation}
s_{12} \simeq \sqrt{{m_d \over m_s}} \,,
\end{equation}
but for the small mixing angles $s_{23}$ and $s_{13}$ a fine-tuning of the
quark phases $\phi_i$ is needed
\begin{equation}
s_{23} \simeq -\sqrt{{m_s \over m_b}} + e^{i\phi_1} \sqrt{{m_c \over m_t}}
\qquad s_{13} \simeq -\sqrt{{m_u \over m_c}} e^{i\phi_2} \cdot s_{23} \,,
\end{equation}
for obtaining the values that have been measured.

The second and more substantial criticism concerns the very existence of any
such relations that could be scale independent. There is namely no apriori
reason why any relation among the various observables of the fermion-mass and
mixing matrices should remain invariant under the renormalization group (RG)
equations which relate them to the structure of the Higgs-Yukawa interaction at
some more fundamental scale. In fact, Olechowski and Pokorski \cite{Pokorski}
have shown that such approximate low-energy relations can be preserved in the
presence of a strong top-Yukawa coupling, or more generally, when the Yukawa
couplings of one fermion family become predominant, a fact that holds true for
the third generation.
This was shown by expressing the masses and mixing angles in terms of
invariants of the Yukawa matrices and then study their evolution, assuming
that the latter is governed by the radiative corrections to the gauge and the
Higgs-Yukawa couplings coming from the MSSM, or other models containing
two doublets of Higgs bosons, or simply the SM.
Interestingly, they obtained some universal results. First, the evolution
of the Cabibbo angle is for all models negligible.
Writing the CKM matrix in the Wolfenstein parametrisation \cite{Wolfenstein}:
\begin{equation}
V_{CKM} = \left(
\begin{array}{ccc}
1 - \lambda^2/2 & \lambda & A \lambda^3 (\rho + i \eta) \\
 - \lambda & 1 - \lambda^2/2 & A \lambda^2 \\
A \lambda^3 (1 - \rho + i \eta) & - A \lambda^2 & 1
\end{array}
\right) \,,
\end{equation}
where the small expansion parameter $\lambda \simeq 0.22$ is approximately the
Cabibbo angle, one finds that only the parameter $A \simeq 0.9$ (and the
CP phase) evolves when going from a low-energy scale $M_0$ to a high-energy
scale $M_X$:
\begin{equation}
{dA \over dx} \simeq - {3 c_0\over 2} (h_t^2 + h_b^2) A \,,
\end{equation}
while:
\begin{equation}
{d\lambda \over dx} \sim {\cal O}(\lambda^4) \,,
\end{equation}
where $x = 1/16 \pi^2\, ln (M_X/M_0)$, $h_t$ and $h_b$ are the top and bottom
Yukawas, and the constant $c_0$ is determined by the gauge couplings of the
model, {\it e.g.} $c_0=2/3$ for the MSSM.
An interesting consequence of this evolution behaviour is that in the MSSM the
small mixing elements $|V_{13}|$ and $|V_{23}|$ become smaller with increasing
energy. This may suggest a mass-generation mechanism where due to some yet
unknown symmetry principle at ultra-high
energies, {\it i.e} the Planck scale, only the fermions of the third generation
are having a mass, but as their mixing with the fermions of the first and
second generation gets stronger at lower energies they too develop a (smaller)
mass.
As for the evolution of the quark-mass ratios the following approximate
equations hold \cite{Pokorski}:
\begin{equation}
{d (m_{u,c}/m_t) \over dx} \simeq - {3 \over 2}
(c_1 h_t^2 + c_0 h_b^2)\, (m_{u,c}/m_t)  \,,
\end{equation}
and
\begin{equation}
{d (m_{d,s}/m_b) \over dx} \simeq - {3 \over 2}
( c_0 h_t^2 + c_1 h_b^2)\, (m_{d,s}/m_b)  \,,
\end{equation}
where $m_{u/c}$ stands for $m_u$ and $m_c$, and
$m_{d/s}$ for $m_d$ and $m_s$, and where for the MSSM $c_1 = 2$.
So in the approximation where only the Yukawa couplings of the third generation
are considered, the different mixing elements run in the same way
as the corresponding fermion-mass ratios and talking about the existence of
such approximate relations makes indeed sense.


\section{Implementing the neutral-lepton sector into the Yukawa quilt.}
While the search for Ans\"atze for the up- and down-quark mass matrices and for
the
charged leptons beyond the electroweak scale needs some extra motivation, for
the neutral-lepton mass matrices, this approach represents a necessity, as they
are not part of the SM. The particular choice of the grand-unification scale as
the scale where such Ans\"atze are formulated
follows of course from the same arguments that motivated the RRR work
\cite{RRR}, namely
the unification of the gauge couplings within the MSSM and some of the Yukawa
couplings in grand-unified theories.
The major difficulty that one is faced with is
the lack of uniqueness in the choice of the three $3\times 3$ Yukawa matrices
$Y_u$, $Y_d$ and $Y_e$ whose diagonalisation should lead to the observable
quark and charged-lepton masses:

\begin{eqnarray}\label{upD}
M_u^{diag} & = & <v_1> U_u^L\, Y_u\, U_u^{R\dag} \nonumber \\
           & = & \left( \begin{array}{ccc}
                     m_u & 0 & 0 \\
                     0 & m_c^{\prime}/\lambda^4 & 0 \\
                     0 & 0 & m_t^{\prime}/\lambda^8
                     \end{array} \right) \,,
\end{eqnarray}

$$m_c^{\prime} \sim m_c \lambda^4 \qquad m_t^{\prime}\sim m_t \lambda^8 \,,$$

\begin{eqnarray}\label{dD}
M_d^{diag} & = & <v_2> U_d^L\, Y_d\, U_d^{R\dag} \nonumber \\
           & = & \left( \begin{array}{ccc}
                     m_d & 0 & 0 \\
                     0 & m_s^{\prime}/\lambda^2 & 0 \\
                     0 & 0 & m_b^{\prime}/\lambda^4
                     \end{array} \right) \,,
\end{eqnarray}

$$m_s^{\prime} \sim m_s \lambda^2 \qquad m_b^{\prime}\sim m_b \lambda^4 \,,$$

and:
\begin{eqnarray}\label{eD}
M_e^{diag} & = & <v_2> U_e^L\, Y_e\, U_e^{R\dag} \nonumber \\
           & = & \left( \begin{array}{ccc}
                     m_e & 0 & 0 \\
                     0 & m_{\mu}^{\prime}/\lambda^4 & 0 \\
                     0 & 0 & m_{\tau}^{\prime}/\lambda^6
                     \end{array} \right) \,,
\end{eqnarray}

$$m_{\mu}^{\prime} \sim m_{\mu} \lambda^4 \qquad m_{\tau}^{\prime}\sim
\noindent
m_{\tau} \lambda^6 \,,$$
where $<v_1>$ and $<v_2>$ are the two vacuum expectation values giving mass to
the up- and down-type fermions. The masses have been parametrised
$\grave{a}\, la$
Wolfenstein \cite{Wolfenstein} so that the order of magnitude of the various
elements becomes manifest.

In order to determine the unitary transformation matrices $U_{u,d,e}$,
anomalous
and hence the structure of the original Higgs-Yukawa interaction sector,
simplifying assumptions are needed. Since the guiding principle has
always been to look for symmetries and to be able to make predictions
the requirement of looking for {\it Ans\"atze} with a maximal number of zeros
compatible with the non-singularity of the $Y_i$'s seems to be a reasonable
approach to this problem \cite{Fritzsch},\cite{DHR},\cite{Giudice},\cite{RRR}.
In order to limit the possible choices any further it has been also assumed
that the Yukawa matrices should be hermitian, so that
$U^L = U^R \equiv U$. This would be the case if the Higgs-Yukawa interactions
were symmetric in generation space.
Analysing all possible choices which satisfy these requirements and are in
agreement with the present experimental data, RRR \cite{RRR} found a set of
five distinct classes, each characterised by a particular structure for $Y_u$
and $Y_d$ at $M_G$. For convenience we write the Yukawa matrix of the up-type
quarks in the following way:
\begin{equation}
Y_u = \left(
\begin{array}{ccc}
0 & \alpha \lambda^6 & \delta \lambda^4 \\
\alpha \lambda^6 & \beta \lambda^4 & \gamma \lambda^2 \\
\delta \lambda^4 & \gamma \lambda^2 & 1
\end{array}
\right) \,,
\end{equation}
where the parameters $\alpha$, $\beta$, $\gamma$ and $\delta$ help to classify
the different cases according to table (1). It is interesting to note that in
the solutions (I), (II) and (IV) found by RRR the up-quark mass matrices
are of the (generalized) Fritzsch-type \cite{Fritzsch},
while in the solutions (III) and (V)
they are of a different type, first proposed by Giudice \cite{Giudice}.
It is basically the parameter $\delta$ that distinguishes between the two  type
of models, being equal to zero in the first case and different from zero in the
latter.
In contrast the down-quark matrices are always of the Fritzsch-type:
\begin{equation}
Y_d = \left(
\begin{array}{rcl}
0 & \alpha^{\prime} \lambda^4 & 0 \\
\alpha^{\prime} \lambda^4 & \beta^{\prime} \lambda^3 & \gamma^{\prime}
\lambda^3 \\
0 & \gamma^{\prime} \lambda^3 & 1
\end{array}
\right) \,,
\end{equation}
with the corresponding values of the parameters $\alpha^{\prime}$,
$\beta^{\prime}$, $\gamma^{\prime}$ shown also in table (1).

Before turning to the lepton sector of these five classes of models
some conceptual clarification is needed. The attentive reader must
have namely noticed that so far there has been no ingredient whatsoever
in the above classification which could justify them being refered to
as GUT models. First they do not represent models but mere classifications
of possible models. Second, as long as the different fermion sectors are
treated as being apriori independent from each other, the idea of grand
unification has not been, strictly speaking, implemented. This will be the case
when we relate the lepton to the quark sector.
Of course the simplest and most natural realisation of the idea of grand
unification would lead to all fermion masses of each generation being equal,
a case already ruled out by experiment.
Therefore in the construction of phenomenologically viable GUT models a
``rich'' Higgs sector is needed in order to differentiate between the up- and
down-quark masses on one side, and between the quark and lepton sectors on the
other side. In the original GUT models based on the $SO(10)$ group this has
been {\it e.g.} achieved by introducing more Higgs fields in the ${\bf 10}$,
${\bf 16}$, the ${\bf 45}$ and the ${\bf 126}$ representations of the group
\cite{oldGUT}.
For some of the superstring-derived or superstring-inspired GUT models, like
those based
on the $SU(4)\times SU(2)_L\times SU(2)_R$ and the flipped $SU(5)\times U(1)$
groups \cite{Antoniadis} - when they are embedded into the $SO(10)$ -, the
absence of Higgs bosons in adjoint or any higher representations became a
problem and alternative mechanisms have been employed \cite{Ranfone}. In any
case, and independently of the chosen path, the most economic way in doing so
is to keep the two up-type Yukawa sectors and the two down-type Yukawa sectors
separately, and up to minor modifications, proportional to each other :
\begin{equation}
Y_u^{ij} \simeq Y_{\nu D}^{ij} \qquad  Y_d^{ij}\simeq Y_e^{ij} \,,
\end{equation}
where $Y_{\nu D}^{ij}$ are the Dirac-type Yukawa couplings of the neutral
lepton sector.
In order that the left equality leads to phenomenologically acceptable
light neutrino masses, there must also exist a
mechanism that generates heavy masses of ${\cal O}(R)$ for the right-handed
singlet states, so that the seesaw mechanism of equ.(\ref{Mnu}) becomes
effective.

These mass terms can come from different sources, - directly from tree level
couplings to Higgs fields or radiatively from loop contributions -, and can a
priori lie in any energy range above a few TeV. In grand-unified models that
have been inspired from the superstrings they have been linked to
nonrenormalisable operators that are abundantly present after string
compactification \cite{nonren}:
\begin{equation}
M_{R_{ij}} = {C\over M_S} {\bar N}^c_{L_i} N_{R_j} <H> <H> \,,
\end{equation}
where $M_S$ is the string unification scale, $H$ a Higgs field in the ${\bf
10}$
representation of $SO(10)$ developing a vacuum expectation value at $M_G$,
and $C \sim e^{-R_s^2/\alpha_s}$ a scale related to the radius  $R_s$ and
the string tension $\alpha_s$ of the Calabi-Yau space.
For values of $C \simeq 10^{-3} - 1$, the entries in the heavy Majorana
matrix will be of the order:
\begin{equation}\label{Planck}
M_{R_{ij}} = (10^{11} - 10^{14}) \, GeV \,.
\end{equation}
In this type of models, but also in general, the structure of $M_R$ is unknown.
Only in special cases where {\it e.g.} the matrices $M_R$ and $M_{u(d)}$ are
proportional due to some constraints, one can make
definite predictions.
Given this fact,
the best way to satisfy the ``principle of minimality'' is to leave $M_R$ as
general as possible, {\it i.e.}, allowing for arbitrary entries $R_{ij}$
$(i,j = 1,2,3)$ as long as the determinant of $M_R$ is nonzero, and disregard
any possible phases. The requirement of no extra phases in the lepton sector
with respect to the quark sector limits the number of free parameters, thus
improving predictibility. To this we shall add one more constraint, namely that
$M_R$ like all the other fermion-mass matrices is symmetric in generation
space, and write it as follows:
\begin{equation}
M_R = \left(
\begin{array}{ccc}
R_1 & R_4 & R_5 \\
R_4 & R_2 & R_6 \\
R_5 & R_6 & R_3
\end{array}
\right) \,.
\end{equation}

If the heavy Majorana mass matrix $M_R$ is not singular, then all the
heavy mass eigenstates of ${\cal O}(R)$ will decouple at energies below their
mass scale, leaving behind (after block diagonalisation) an effective mass
matrix for the light Majorana states:
\begin{equation}\label{EM}
M_{\nu}^{eff} \simeq M_u M_R^{-1} M_u^{\dag} \,.
\end{equation}

Denoting the inverse of the heavy Majorana matrix as:
\begin{equation}\label{MR1}
M_{R_{ij}}^{-1} = {r_{ij}\over \Delta}
\end{equation}
with:
\begin{equation}
\begin{array}{lll}
r_1 \equiv r_{11} = R_2 R_3 - R_6^2 & &
r_4 \equiv r_{12} = r_{21} = R_5 R_6 - R_3 R_4 \\
r_2 \equiv r_{22} = R_1 R_3 - R_5^2 & &
r_5 \equiv r_{13} = r_{31} = R_4 R_6 - R_2 R_5  \\
r_3 \equiv r_{33} = R_1 R_2 - R_4^2 &
& r_6 \equiv r_{23} = r_{32} = R_4 R_5 - R_1 R_6 \,,\\
\end{array}
\end{equation}
and $\Delta = det M_R$, the effective light-neutrino mass matrix is given by:
\begin{equation}
M_{\nu_{ij}}^{eff} = {m_t^2 \over \Delta}\, m_{ij} \,.
\end{equation}
with:
\begin{equation}
\begin{array}{l}
m_{11} = \delta^2 r_3 z^4 + \alpha^2 r_2 z^6 \\
m_{12} = m_{21} =  \gamma \delta r_3 z^3 + (\beta \delta + \alpha \gamma) r_6
z^4 + \alpha \beta r_2 z^5 + \alpha^2 r_4 z^6 \\
m_{13} = m_{31} = \delta r_3 z^2 + (\alpha + \gamma \delta) r_6 z^3
          + (\alpha \gamma r_2 + \delta^2 r_5) z^4 \\
m_{22}  =  \gamma ^2 r_3 z^2 + 2 \beta \gamma r_6 z^3 + \beta^2 r_2 z^4
           + 2 \alpha \gamma r_5 z^4 + 2 \alpha \beta r_4 z^5
           + \alpha^2 r_1 z^6 \\
m_{23}  = m_{32} = \gamma r_3 z + (\gamma^2 + \beta) r_6 z^2
           +(\beta \gamma r_2 + \alpha r_5 + \gamma \delta r_5) z^3
           +(\beta \delta + \alpha \gamma) r_4 z^4 \\
m_{33}  =  r_3 + 2 \gamma r_6 z + (\gamma^2 r_2 + 2 \delta r_5) z^2
           + 2 \gamma \delta r_4 z^3 + \delta^2 r_1 z^4 \,,
\end{array}
\end{equation}
and where we have set $\lambda^2 = z \simeq 0.05$.
The matrix elements of $M_{\nu}^{eff}$ are polynomials in the small parameter
$z$ with the minors of the matrix $M_R$, the $r_{ij}$'s, as coefficients.

We can now start with the discussion on possible classification schemes of the
neutrino sector of maximally predictive GUT models. The first thing to notice
is that when
\begin{equation}
 r_3 \not= 0 \qquad {\rm and} \qquad r_3 \geq r_{ij} \,,
\end{equation}
it sets, independently of the model, the scale for $m_{33}$, which being of
zero power in $z$ is anyway the largest entry, and for the entire matrix.
This condition is in particular satisfied when all the entries of the matrix
$M_R$ are of the same order of magnitude and there are no cancellations among
them so that none of its invariants becomes singular.

The first case {\bf (i)} that we will therefore consider is defined through the
conditions:
\begin{equation}\label{C1}
\begin{array}{c}
M_{R_{ij}} \sim \cal{O}(R) \\
r_{ij} \not=  0 \,, \\
\end{array}
\end{equation}
which imply that the heavy Majorana sector contains one mass scale only and no
extra symmetries.
Under these two assumptions the overall scale of the effective light-neutrino
matrix is given approximately by:
\begin{equation}\label{m0}
m_0 = m_t^2\, {r_3\over \Delta} \, \simeq \, {m_t^2\over R}  \,.
\end{equation}
The structure of $M_{\nu}^{eff}$ to lowest order in $z$ is shown for the five
different classes of maximally predictive GUT models (from ref.(\cite{RRR}))
in table (2) and is written in terms of the ratios:
\begin{equation}
a_1 \equiv {r_2\over r_3}\quad a_2 \equiv {r_6\over r_3}\quad  a_3 \equiv
{r_4\over r_3} \quad
a_4 \equiv {r_5\over r_3} \quad a_5 \equiv {r_1\over r_3} \,.
\end{equation}
There are two things that one notices immediately. First, there
is a strong hierarchy of the Fritzsch-type among the entries of the second and
third generation ($m_{22}\sim m_{23}^2 \ll m_{33}$) in all five classes, given
by $\lambda^2$ or even by $\lambda^4$.
As far as the mixing with the first generation is concerned, there is a
breaking of the usual pattern of only nearest-neighbour generation couplings,
known from the quark sector, where in particular
$m_{13}\leq m_{22}$. Depending on the class, the $m_{13}$ element can be larger
than $m_{22}$ (classes (I) and (III)), while it is always larger than $m_{12}$.
This is an interesting example of a nontrivial case where the heavy Majorana
matrix was not chosen to be proportional to the up-quark mass matrix so that
one does also expect the proportionality between the mixing matrices
$V_l$ and $V_{CKM}$ to be broken, a situation leading to interesting
phenomenological consequences. For the same reason a small $m_{11}$ entry
appears in the effective light-neutrino matrix .

The structure of the effective neutrino matrix is significantly altered
in some cases when either of the two conditions given by equ.(\ref{C1})
or both are not satisfied. We study the different possibilities separately, and
start with the case where some of the $r_i$'s are zero,
except for $r_3$, meaning that the matrix $M_R$ has some extra symmetries.
In this case {\bf (ii)}, defined through:
\begin{eqnarray}\label{C2}
M_{R_{ij}} & \sim & {\cal O}(R) \nonumber \\
r_i & = & 0 \qquad i\subset \{1,2,4,5,6\} \,,
\end{eqnarray}
the hierarchy among the entries of $M_{\nu}^{eff}$ may be altered with respect
to the previous case, but not as drastically as to lead to a complete reversal
of the natural-hierarchy pattern.

The situation may differ for the case {\bf (iii)}:
\begin{eqnarray}\label{C3}
{R_j \over R_k} & \simeq & {\cal O}({\bf z^n}) \nonumber \\
{r_i \over r_l} & \simeq & {\cal O}({\bf 1}) - {\cal O}({\bf z^{2n}}) \,,
\end{eqnarray}
where the matrix $M_R$ is having more scales, characterised by a strong
hierarchy, so that some of the $r_i$ minors can be considerably enhanced with
respect to others, thus changing completely the structure of $M_{\nu}^{eff}$
- to leading order in z - that is shown in table (2).
In some special cases the distortion of the perturbative structure
of $M_{\nu}^{eff}$ could go as far as to render all the matrix entries
comparable to each other, or even reverse their hierarchical order.
This possibility is limited though by the fact that the same $r_{ij}$
minors enter as coefficients to different powers of $z$ in the various
$m_{ij}$ elements, thus protecting to a certain degree the inate hierarchy
of the matrices. Another interesting possibility is the enhancement of $r_3$
and the matrix scale from $m_0$ to:
\begin{equation}
m_0^{\prime} = p \times m_0
\end{equation}
by a factor $p$ considerably larger than one, so that,
if $m_0^{\prime}$ would be the scale of the heaviest neutrino as the HDM
candidate, that would correspond to a scale:
\begin{equation}
R^{\prime} \simeq  p \times 10^{12} \, GeV \,,
\end{equation}
that can nicely fit within the range expected from residual nonrenormalisable
terms from the Planck scale, equ.(\ref{Planck}).

Finally the last and most interesting case {\bf (iv)} arises when:
\begin{equation}\label{C4}
r_3 = 0 \,,
\end{equation}
because this may imply a very light tau neutrino and therefore a realisation of
the inverse hierarchy scenario of eqs.(\ref{RH1},\ref{RH2}).
This is in particular the case for the
first RRR class of models for which $m_{33} = 0$, while $m_{23} \sim {\cal
O}(z^2)$ and $m_{22} \sim {\cal O}(z^3)$.
Another case of this type arises for models belonging to class (III), because
the leading order behaviour of the corresponding $m_{ij}$ entries contains
different $r_{ij}$ coefficients. For example, if $r_5 = 0$ (and eventually
also $r_4 = r_1 = 0$), but $r_6$ or $r_2$ are nonzero, then $m_{23} \gg
m_{33}$. Notice that for models belonging to any of the other RRR classes (I),
(II), (IV) and (V),  an ``anomalous'' ordering: $m_{33}\ll m_{23},m_{22}$
is not possible as long as $m_{33} \not= 0$.
On the other hand, for all classes except for class (V), $m_{33}$ can, due to
extra symmetries in $M_R$, be zero without all the other entries being
necessarily zero. The breaking of the natural ordering among the generations
appears as an interesting possibility, common to four out of five classes of
maximally-predictive GUT models, that will be discussed separately.
In contrast, for those cases where $m_{33}$ maintains its role as the
predominant entry, the overall scale of the matrix is reduced from $m_0$ to:
\begin{equation}
m_0^{\star} = m_t^2 \, {r_{ij} z^n \over \Delta}
\simeq {m_t^2 \over R} \,z^n  \,,
\end{equation}
where $r_{ij}$ and the power of $z$ are model dependent, and the second
equality holds in the special case of no hierachy in $M_R$. This obviously
modifies the range where $R$ should lie if $m_0^{\star}$ were to be the scale
of the heaviest neutrino that would correspond to a HDM candidate. Instead of
equ.(\ref{R}) one would then obtain a smaller intermediate scale:
\begin{equation}\label{RP}
R^{\star} \sim z^n \times 10^{12}\, GeV \,.
\end{equation}

\section{The spectrum of light neutrinos in different classes of
maximally-predictive  grand-unified models.}
After the classification of the heavy Majorana sector into four classes,
cases ({\bf i - iv}), with respect to the
structure of the effective light-neutrino mass matrix, we turn to the
determination of the masses and mixings of the latter.
The mass eigenvalues of the matrix $M_{\nu}^{eff}$ can be found from a
perturbative determination of the roots of its characteristic polynomial,
written as:
\begin{equation}\label{P}
P = x^3 - r_3 f x^2 + z^4 r_1^{\star} g x - \Delta_{\nu} \,,
\end{equation}
where by $\Delta_{\nu}$ we have denoted the determinant of $M_{\nu}^{eff}$, by
$r_1^{\star} = r_2 r_3 - r_6^2$ the corresponding minor of the matrix $r_{ij}$
and $f$ and $g$ are polynomials in $z$:
\begin{eqnarray}\label{g}
f & = & 1 + 2 \gamma a_2 z + [(1 + a_1) \gamma^2 + 2 \delta a_4] z^2
        + 2 \gamma (\delta a_3 + \beta a_2) z^3 \nonumber \\
  &   & + [(1 + a_5) \delta^2 + \beta^2 a_1 + 2 \alpha \gamma a_4] z^4
        + 2 \alpha (\beta a_3 + \delta a_2) z^5 \nonumber \\
  &   &  + (a_1 + a_5) \alpha^2 z^6 \\
g & = & 1 - 2 a_1^{\star}
    (\alpha \gamma^2 - \alpha \beta + \beta \gamma \delta - 2 \gamma^3) z
\nonumber \\
  &   & + {\cal O}(z^2) + ... + {\cal O}(z^8) \nonumber \,,
\end{eqnarray}
with $a_1^{\star} = (r_3 r_4 - r_5 r_6) /r_1^{\star}$.


\subsection{Case {\bf (i)}: Neutrino spectra with a ``natural'' mass
hierarchy.}
We start our discussion with the first case {\bf (i)} of a single scale $R$ and
no symmetry for $M_R$ so that $f$ and $g$ are now functions of order one.
Redefining next
$x \to x/R^2$ and using the fact that:
\begin{equation}
\begin{array}{c}
\Delta_{\nu} \simeq \kappa^2 z^{12} \cdot {\rm det} M_R^{-1} \\
\kappa = \alpha^2 + \beta \delta^2 - 2 \alpha \gamma \delta \simeq {\cal O}
({\bf 1})\,,
\end{array}
\end{equation}
equ.(\ref{P}) reduces to:
\begin{equation}
P_0 \simeq x^3 - x^2 + z^4 x - z^{12} \,,
\end{equation}
which is the same for all five RRR classes of models. Therefore under the
minimality condition of no hierarchy and no symmetry for the heavy Majorana
sector, one obtains an entirely model-independent neutrino-mass spectrum:
\begin{eqnarray}\label{SP0}
m_{\nu_1} & \simeq & {m_t^2\over R} z^8 \nonumber\\
m_{\nu_2} & \simeq & {m_t^2\over R} z^4 \\
m_{\nu_3} & \simeq & {m_t^2\over R} \nonumber\,.
\end{eqnarray}
The hierarchy implied by eqs.(\ref{SP0}) is of the quadratic-seesaw type:
\begin{eqnarray}\label{H0}
{m_{\nu_1}\over m_{\nu_2}} & \simeq & {m_u^2\over m_c^2} \simeq z^4 \\
{m_{\nu_2}\over m_{\nu_3}} & \simeq & {m_c^2\over m_t^2} \simeq z^4\,.
\end{eqnarray}
This result strictly holds at the unification scale $M_G$ and approximately at
the scale $M_X$ at which the heavy states decouple.
For deriving the right-hand side we have used the fact that the running of the
quark-mass ratios from the electroweak scale $M_Z$ to $M_G$, when the
top-Yukawa coupling is assumed to be constant and threshold effects are
neglected, is controlled by the parameter
$\chi = (M_G/M_Z)^{-h_t^2/16\pi^2} \simeq 0.7$ \cite{RRR} :
\begin{eqnarray}
{m_{u,c}\over m_t} (M_G) & \simeq & \chi^3 {m_{u,c}\over m_t} (M_Z) \\
{m_{d,s}\over m_b} (M_G) & \simeq & \chi {m_{d,s}\over m_b} (M_Z)\,.
\end{eqnarray}

We turn next to the determination of the lepton-mixing matrix $V_l$.
Motivated by the successes of the GJ Ansatz, the texture structure of the
charged-lepton Yukawa matrix $Y_e$ at $M_G$ will be chosen to be the same as
for $Y_d$ except for the (2,2) entry which will be multiplied by a factor of
minus three. For the same minimality reasons we mentioned before we will assume
no extra CP-violating phases.
Defining by:
\begin{equation}
U_P = \left(
\begin{array}{ccc}
1 & 0 & 0 \\
0 & 1 & 0 \\
0 & 0 & e^{i\phi}
\end{array}
\right) \qquad ,
\end{equation}
the matrix relating the basis where $M_{\nu}^{eff}$ is diagonal to the basis
where $M_e$ is real,
\begin{equation}
V_l = U_{\nu} \, U_P \, U_e^{-1} \,,
\end{equation}
where $U_{\nu}$ and $U_e$ are the matrices diagonalising $M_{\nu}^{eff}$
and $M_e$ respectively. Written in powers of $\lambda$ and to lowest order,
\begin{equation}
U_e = \left(
\begin{array}{lcr}
1-\lambda^2/18 & -\lambda/3 & \gamma \lambda^4/3 \\
\lambda/3 & 1-\lambda^2/18 & -\gamma\lambda^3 \\
0 & \gamma \lambda^3 & 1
\end{array}
\right) \qquad .
\end{equation}
Except for class (V) the structure of the lepton-mixing matrix, again under the
assumption of no hierarchy and no symmetry for the heavy Majorana sector, is
universal:
First, the mixing between the first two generations is always:
\begin{equation}
\mid V_{\nu_1 - \mu}^{(I-IV)}\mid\, \simeq\, s^l_{12}\, \simeq \, {\lambda\over
3}\, \simeq \, 0.07 \,,
\end{equation}
where the factor-three reduction with respect to the Cabibbo angle is a direct
consequence of the GJ relation.
This leads to a $\nu_e - \nu_{\mu}$ mixing of:
\begin{equation}
sin^2 2\theta_{e - \mu}\, \simeq \, 0.02 \,,
\end{equation}
which falls naturally within the range required by the small-angle MSW solution
to the solar neutrino problem, equ.(\ref{MSW}).
In contrast, in models of the class-(V) type the mixing is two to three times
too large.
\begin{equation}
sin^2 2\theta_{e - \mu}\, \simeq\, 0.05 \,.
\end{equation}

The mixing between the second and third generation is:
\begin{equation}
\mid V_{{\nu_2} - \tau}^{(I,III)}\mid\, \simeq\, s^l_{23} \,\simeq\, 4
\lambda^3 \,\simeq\, 0.03
\end{equation}
for models belonging to the classes (I) and (III), giving rise
to a $\nu_{\mu} - \nu_{\tau}$ mixing angle:
\begin{equation}
sin^2 2\theta_{\mu - \tau}\, \simeq\, 7. \times 10^{-3} \,,
\end{equation}
while it is somewhat larger in models of the type (II) and (IV):
\begin{equation}
\mid V_{{\nu_2} - \tau}^{(II,IV)}\mid \, \simeq \, \lambda^2 \qquad
sin^2 2\theta_{\mu - \tau}\, \simeq \, 9. \times 10^{-3} \,.
\end{equation}
For the anomalous case (V) the mixing between $\nu_{\mu}$ and $\nu_{\tau}$
is negligible.
Finally the mixing between the first and third generation is:
\begin{equation}
\mid V_{\nu_1 - \tau}^{(I,II,IV)}\mid \,\simeq\, s^l_{13}\, \simeq\, \alpha
\lambda^6
\end{equation}
for models of the type (I), (II) and (IV), giving rise
to a $\nu_e - \nu_{\tau}$ mixing angle:
\begin{equation}
sin^2 2\theta_{e - \tau} \,\simeq\, (0.5 - 1.) \times 10^{-7} \,.
\end{equation}
In models of the type (III) and in particular of the type (V) the
first-to-third generation mixing is considerably enhanced:
\begin{eqnarray}
\mid V_{\nu_1 - \tau}^{(III)}\mid & \simeq & \lambda^4
\qquad sin^2 2\theta_{e - \tau} \simeq 2.2 \times 10^{-5} \\
\mid V_{\nu_1 - \tau}^{(V)}\mid & \simeq & \lambda^2
\qquad sin^2 2\theta_{e - \tau} \simeq 9. \times 10^{-3} \,.
\end{eqnarray}
For the last type of models the value of the $\nu_e - \nu_{\tau}$
mixing angle is rather representative of what one would expect for $\nu_{\mu} -
\nu_{\tau}$ mixing.

One can summarise these results by saying that apart from a certain variation
in the values of the lepton-mixing angles, the four classes (I - IV) of
GUT models, in the presence of a single scale $R$ and the absence of any extra
$M_R$ symmetries, lead to the following universal mixing pattern:
\begin{equation}
\mid V_{\nu_1 - \mu}\mid \,\sim\, {\lambda\over 3}\, \gg \,
\mid V_{\nu_2 - \tau} \mid\, \sim\, \lambda^2 - 4\lambda^3 \, \gg\,
\mid V_{\nu_1 - \tau} \mid \,\sim\, \lambda^4 - \lambda^6 \,.
\end{equation}
The numerical range of the mixing angles is close to the naive seesaw-based
estimates in section 2.2 , resulting from a scenario that incorporates the
small-angle solution to the solar neutrino problem and the tau neutrino as a
candidate for the HDM. They confirm the simple guess that:
\begin{equation}
s^l_{12} \,<\, s_{12} \qquad s_{23}^l \,<\, s_{23} \qquad s^l_{13}
\,\ll\,s_{13} \,.
\end{equation}
However, due to the running of the up-quark masses, the hierarchy of the light
neutrino masses at $M_G$ is best described by the quadratic seesaw rather than
the low energy mixed-seesaw relations, eqs.(24 - 26).

Only the ``anomalous'' class (V) models break this pattern.
Due to the predictions of a vanishing $\nu_{\mu} - \nu_{\tau}$ mixing and a
rather large $\nu_e - \nu_{\tau}$ mixing they represent a quite distinct case.

\subsection{Case ({\bf ii}): Neutrino spectra with a slightly distorted mass
hierarchy.}

In the preceding section we have examined classes of maximally predictive GUT
models with the least number of constraints imposed upon the structure of the
heavy-neutrino Majorana mass matrix, namely the case of no hierarchy of
scales and no underlying symmetry principle, and we have obtained a universal
spectrum of masses and mixing angles for the three light neutrinos. Their
hierarchy was considered as the most natural since it corresponds to what
was expected from the simplest seesaw scenario.
In this and the following section we shall discuss classes of models with a
more elaborate structure in what concerns the heavy Majorana-mass sector which
give a distorted neutrino spectrum with respect to the previous case (i). It is
to be expected that
such cases will arise in the presence of a strong (inverse) hierarchy of large
mass scales and/or in the presence of extra symmetries.

We will first focus on the possibility that as a result of such symmetries
some of the subdeterminants of $M_R$ are zero and therefore certain powers of
$\lambda$ in the effective light-neutrino mass matrix are suppressed.
We start with the case {\bf (ii)} where there is a single scale $R$ and
$r_3 \not= 0$, but allow some of the ratios $a_i$ to be zero rather than of
order one, equ.(\ref{C2}). The characteristic polynomial of the matrix
$M_{\nu}^{eff}$ assumes to leading order now the more general form:
\begin{equation}
P_1 \simeq x^3 - x^2 + x z^n - z^{12} \,,
\end{equation}
where $n$ is an integer between four and six. Though the overall scale of
$M_{\nu}^{eff}$, set by the heaviest state, the $\nu_{\tau}$, remains
unchanged, as given by equ.(\ref{m0}), the splitting between the mass
eigenstates can be in general different from the case studied previously:
\begin{equation}
{m_{\nu_2}\over m_{\nu_3}} \simeq z^n \qquad
{m_{\nu_1}\over m_{\nu_3}} \simeq z^{12-n} \qquad n = 4,5,6 \qquad,
\end{equation}
depending upon the leading power of the polynomial $g$, equ.(\ref{g}).
In addition to the spectrum of equ.(\ref{SP0}) which corresponds to the case
$n = 4$, there are namely two more neutrino-mass spectra:
\begin{equation}
m_1 : m_2 : m_3 = \left\{
\begin{array}{ccccc}
 z^7 & : & z^5 & : & 1 \\
 z^6 & : & z^6 & : & 1 \,,
\end{array}
\right\} \times m_0 \,,
\end{equation}
where the first exhibits a reduced hierarchy with respect to the natural
hierarchy and the second an approximate mass degeneracy among the neutrinos of
the first- and second-generation.
In table (3) we show possible values for the lepton-mixing angles in the
five type of models considered before, setting the parameters $a_i$
subsequently equal to zero.
We have limited ourselves to those cases, where the perturbative structure of
$M_{\nu}^{eff}$ is considerably altered with respect to table (2).
In general, we have not found any substantial distortion in the spectra as
compared to case {\bf (i)}, except for some special cases where the mixing
angle between the first and third generation or the second and third generation
are zero. Otherwise, one finds also for the case {\bf (ii)} a
universal lepton-mixing pattern:
\begin{equation}
\mid V_{\nu_e - \mu}\mid \,\sim\, {\lambda\over 3}\, \gg \,
\mid V_{\nu_{\mu} - \tau}\mid \,\sim\, \lambda^3 \, \gg
\mid V_{\nu_e - \tau}\mid \,\sim\, \lambda^3 - \lambda^5 \,,
\end{equation}
which qualitatively resembles the one of case {\bf (i)}.
For the models belonging to class (V), for which $r_3 \not= 0$
enters as a coefficient in all $m_{ij}$ elements to leading order, the
resulting spectrum is basically the same as in the previous case {\bf (i)}.


\subsection{Case (iii): Neutrino spectra in the presence of a strong hierarchy
of large Majorana-mass scales.}

We consider next the case {\bf (iii)} where, due to a strong hierarchy among
the entries of heavy Majorana matrix $M_R$, there is an even stronger splitting
among the $r_{ij}$ coefficients in $M_{\nu}^{eff}$, equ.(\ref{C3}). If $p$ and
$q$ are enhancement (suppression) factors resulting from such splittings in the
polynomials $f$ and $g$, the characteristic polynomial is given by:
\begin{equation}
P_2 \simeq x^3 - p x^2 + q z^4 x - z^{12} \,.
\end{equation}
Then, for $q^2 \geq 4 z^4 p$, one obtains the following neutrino mass
eigenstates:
\begin{equation}
m_{\nu_1}\,:\,m_{\nu_2}\,:\,m_{\nu_3} = ({z^8\over q}\,:\,{q\over p} z^4\,:\,
p) \times m_0 \,,
\end{equation}
that explicitly reflect the distortion of the quadratic seesaw spectrum, found
in case ({\bf i}).
Notice that, for certain values of $p$ and $q$, two of the neutrino
states can become mass degenerate: $m_{\nu_1} \simeq m_{\nu_3}$ or
$m_{\nu_2} \simeq m_{\nu_3}$, while, when $p$ and $q$ are both zero,
all three neutrinos would be mass degenerate.

A detailed analysis of the lepton-mixing sector is complicated, by the
existence of a too large number of possible hierarchical orderings of the
entries of $M_R$. We will therefore adopt a more schematic approach and
concentrate first upon the question of the breaking of the natural ordering for
the second and third generation only.
Let us consider the following types of a mass-matrix structure that arise
naturally in models belonging to classes (I - IV):
\begin{equation}
\begin{array}{cc}
M_0 = \left( \begin{array}{cc}
              z^m & z^n \\
              z^n & 1 \\
             \end{array} \right)
&
M_1 = \left( \begin{array}{cc}
              z^m & 1 \\
              1 & 1 \\
             \end{array} \right)
\\
& \\
& \\
M_2 = \left( \begin{array}{cc}
              1 & z^n \\
              z^n & 1 \\
             \end{array} \right)
&
M_3 = \left( \begin{array}{cc}
              1 & 1 \\
              1 & 1 \\
             \end{array} \right)
\\
\end{array} \,.
\end{equation}
One can easily check that, while for a mass matrix of the type $M_0$
the mixing between the two neutrino states:
\begin{equation}
\nu_2 \,\simeq\, {\rm cos}\theta \nu_{\mu} - {\rm sin}\theta \nu_{\tau}
\qquad
\nu_3 \,\simeq\, {\rm sin}\theta \nu_{\mu} + {\rm cos}\theta \nu_{\tau}
\,,
\end{equation}
is small, sin$\theta_{(0)} \sim z^n$, for matrices of the types $M_1$, $M_2$
and $M_3$ it is maximal, sin$\theta_{(1,2,3)} \sim 2^{-1/2}$, as this would be
required for a solution to the atmospheric neutrino deficit, if it would
persist.

For models belonging to class (I), a large $\nu_{\mu} - \nu_{\tau}$ mixing
can also come from a matrix structure of the type:
\begin{equation}
M_4 = \left( \begin{array}{ccc}
              0 & 0 & 1 \\
              0 & z^n & 1/z \\
              1 & 1/z & 1 \\
             \end{array} \right) \,,
\end{equation}
or in models belonging to class (III) from:
\begin{equation}
M_5 = \left( \begin{array}{ccc}
             0 & z^2 & z^2 \\
             z^2 & 0 & 1 \\
             z^2 & 1 & 1\\
             \end{array} \right) \,.
\end{equation}

One finds also cases leading to large $\nu_e - \nu_{\tau}$ mixing,
like {\it e.g} for models belonging to classes (II) and (IV) having a
mass structure of the type:
\begin{equation}
M_6 = \left( \begin{array}{ccc}
             0 & z & 1 \\
             z & z^2 & z \\
             1 & z & 1\\
             \end{array} \right) \,.
\end{equation}
These would be good candidates for a scenario that could explain the solar
neutrino deficit via the large-angle vacuum-oscillation or MSW solution. In
order to also implement a solution to the HDM problem into this type of
scenario the muon neutrino should be the heaviest state.
Such cases of a completely reversed hierarchy will be discussed next.


\subsection{Case (iv): Neutrino spectra with an inverse mass hierarchy.}

As mentioned previously, an important role in our classification scheme
is attributed to the $r_3$ subdeterminant of $M_R$. When it is zero the leading
order behaviour of the effective light-neutrino mass matrix changes in all five
classes of models. It is zero when, irrespective of the other entries and of
fine-tuning, one has one of the following textures:
\begin{equation}
M_R = \left( \begin{array}{ccc}
             0 & 0 & \star \\
             0 & R_2 & \star \\
             \star & \star & \star\\
             \end{array} \right)
 \qquad {\rm or} \qquad
M_R = \left( \begin{array}{ccc}
             R_1 & 0 & \star \\
             0 & 0 & \star \\
             \star & \star & \star\\
             \end{array} \right) \,.
\end{equation}
As a result, the $(3,3)$ entry of $M_{\nu}^{eff}$ will be zero ({\it e.g} in
models of class (I)) or suppressed by some power of $z$. Let us concentrate
on the first possibility and assume the more general situation where the other
entries can, but need not, be zero. Let us further assume an ``anomalous''
ordering $m_{33} \leq m_{23}, m_{22}$ that is possible for any model belonging
to classes (I - IV), and for simplicity concentrate on those cases with a
zero-mass neutrino state, being some linear combination of $\nu_{\tau}$ with
the other two neutrino flavours. This is then equivalent to the matrix
$M_{\nu}^{eff}$ being singular. Requiring this to be achieved without fine
tuning and through a small number of texture zeros one is led to the following
textures for $M_{\nu}^{eff}$:
\begin{equation}
\begin{array}{ccc}
L^{(4)}_1 = \left( \begin{array}{ccc}
              0 & \star & 0 \\
              \star & \star & \star \\
              0 & \star & 0 \\
             \end{array} \right)
&
L^{(4)}_2 = \left( \begin{array}{ccc}
              \star & \star & \star \\
              \star & 0 & 0 \\
              \star & 0 & 0 \\
             \end{array} \right) \,,
\\
& \\
& \\
L^{(5)}_3 = \left( \begin{array}{ccc}
              0 & 0 & \star \\
              0 & 0 & \star \\
              \star & \star & 0 \\
             \end{array} \right)
&
L^{(5)}_4 = \left( \begin{array}{ccc}
              \star & \star & 0 \\
              \star & \star & 0 \\
              0  & 0 & 0 \\
             \end{array} \right)
\\
& \\
& \\
L^{(5)}_5 = \left( \begin{array}{ccc}
              0 & \star & 0 \\
              \star & 0 & \star \\
              0 & \star & 0 \\
             \end{array} \right)
&
L^{(5)}_6 = \left( \begin{array}{ccc}
             0 & \star & \star \\
              \star & 0 & 0 \\
              \star & 0 & 0 \\
             \end{array} \right)
\\
\end{array} \,,
\end{equation}
with four or five texture zeros respectively.
The star symbol stands for an entry of order one or some power of $z$, implying
strong mixing or weak mixing. Notice first that the texture $L^{(5)}_4$ will
always give a zero mass tau neutrino that does not mix with the other species,
and a more or less strong mixing between the electron and muon neutrinos:
\begin{equation}\label{IS1}
\nu_1 \, =\, \nu_{\tau} \qquad
\nu_2 \, = \, {\rm cos}\theta \nu_e - {\rm sin}\theta \nu_{\mu} \qquad
\nu_3 \, = \, {\rm sin}\theta \nu_e + {\rm cos}\theta \nu_{\mu}
\,.
\end{equation}
On the other hand, the two block diagonal textures $L^{(4)}_1$
and $L^{(5)}_5$ will give a zero-mass eigenstate $\nu_1$ that is a linear
combination of the electron and tau neutrinos, and a muon neutrino (with a
small
admixture of the other two components) as the heaviest state.
\begin{equation}\label{IS2}
\nu_1 \, \simeq \, {\rm cos}\theta \nu_{\tau} - {\rm sin}\theta \nu_e
\qquad
\nu_2 \, \sim \, {\rm sin}\theta \nu_e + {\rm cos}\theta \nu_{\tau}
\qquad
\nu_3 \sim m_{\nu_{\mu}} \,.
\end{equation}
The spectra of eqs.(\ref{IS1},\ref{IS2}) are examples of an inverse
mass hierarchy in the neutrino sector, eqs.(\ref{RH1},\ref{RH2}).
Of the remaining three matrix textures, $L^{(4)}_2$ and $L^{(5)}_6$ will
be never encountered, and $L^{(5)}_3$
will give a zero-mass eigenstate that can be
a linear combination of all three neutrinos with varying degrees of mixing.


\subsection{Radiative corrections}
All our results on neutrino masses and mixing were obtained by assuming an
exact tree-level proportionality between the quark and lepton Yukawa matrices
at the grand unification scale $M_G$ and by subsequently diagonalising the
latter. The resulting mass eigenstates and mixing angles need therefore to be
corrected before any attempt to relate them to the low-energy observables.
As is well known, corrections to the fermion masses come from two sources,
the gauge and the Higgs-Yukawa couplings.
Gauge corrections can be applied to individual eigenvalues, because apart from
differences in mass thresholds they are ``family-blind''. Higgs corrections are
proportional to the Yukawa couplings, but are practically negligible except for
the couplings of the third generation \cite{Pokorski}.
However the treatment of the running of the light neutrino masses from $M_G$ or
$M_X$ down to $M_Z$ or even to the much lower energy scale $M_0$ where
experiments hope to measure them, has been somewhat ambiguous.
Some of the authors \cite{Ellis} have chosen to treat ratios of light neutrino
masses simply as ratios of up-quark masses and consider radiative corrections
only to the latter. Others \cite{Babu} \cite{Shafi} decided not to consider any
radiative corrections to the neutrino masses and mixings, most likely because
they found themselves faced with the problem of a conflicting evolution
behaviour of the $M_u$ and $M_R$ parts of the full neutrino mass matrix. One is
namely
faced with the peculiarity of the seesaw that only part of the physical
spectrum will
give measurable effects at low energies and therefore receive radiative
corrections, while the other part effectively decouples already at the scale
$M_X$. The most reasonable approach seems therefore to consist in calculating
radiative corrections only to the eigenstates of the effective light-neutrino
matrix $M_{\nu}^{eff}$. It is then obvious that this should not be translated
into treating only the nominator of the seesaw masses, since there are no
physical up-quark mass eigenstates contained in a neutrino. In particular there
is no Yukawa coupling for the light neutrinos at the tree level; this is
generated only through radiative corrections. We will therefore adopt the
approach used in ref.(\cite{DHRN}) and complement the evolution equations for
the mass ratios and mixing elements, given by Olechowski and Pokorski
\cite{Pokorski} for the quarks, to include also the leptons. Following the
notation of the latter we write the one-loop RGE for the Yukawa matrices $Y_A$,
where $A = e(\nu)$ stands for the charged leptons (neutrinos), as follows:
\begin{equation}
{d \over dx} Y_A = (c_A {\bf 1} + \sum_B a_{AB} H_B) Y_A \,,
\end{equation}
where $H_A = Y_A Y_A^{\dag}$ is a hermitian matrix, and
$x = (1/16 \pi^2) ln (M_X/M_0)$. In the MSSM the radiative corrections to the
gauge couplings $c_A = G_A - T_A$ are:
\begin{equation}
G_e = 3 g_2^2 + {9 \over 5} g_1^2 \qquad T_e = Tr H_e + Tr H_d
\end{equation}
for the charged leptons and:
\begin{equation}
G_{\nu} = 3 g_2^2 + {3 \over 5} g_1^2 \qquad T_{\nu} = Tr H_u
\end{equation}
for the neutrinos, where $g_1$ and $g_2$ are the electroweak gauge couplings.
The corrections to the Yukawa couplings are specified by $a_{ee} = - 3$ and
$a_{e \nu} = 0$ for the charged leptons, and
$a_{\nu e} = - 1$, $a_{\nu\nu} = 0$ for the neutrinos.

Using the weak-basis invariants and their relations to the mass and mixing
observables, introduced in ref.(\cite{Pokorski}) and based upon the usual
requirements of unitarity and hermiticity and the presence of hierarchy,  which
remain valid also for the lepton Yukawa matrices
\footnote{Due to the decoupling of the heavy Majorana states, $M_{\nu}^{eff}$
is only approximately a unitary matrix, but as long as the mixing of the light
states with the heavy states remains negligible, this does not present any
problem.},
we obtain the following evolution for the Yukawa mass ratios and the mixing
elements of the latter:
\begin{eqnarray}\label{hi}
{d \over dx}\, ln ({h_{e_i}\over h_{e_j}}) & = & 3\, ( h_{e_j}^2 - h_{e_i}^2 )
\simeq 3\, h_{\tau}^2 \\
{d \over dx}\, ln ({h_{\nu_i}\over h_{\nu_j}}) & = &
- \sum_{k=1}^3 \, h_{e_k}^2 \, ( \mid V_{ik}^l\mid^2 - \mid V_{jk}^l\mid^2 ) \\
 & \simeq & - h_{\tau}^2 \, (\mid V_{i3}^l\mid^2 - \mid V_{j3}^l\mid^2 ) \\
 & \simeq &  h_{\tau}^2  \,.
\end{eqnarray}
In going from the first to
 the third  line of the last equation we have made use of
the predominance of the third generation Yukawa coupling and the fact that the
$\mid V_{\nu_3\tau}\mid$ mixing element is of order one. Neglecting the small
corrections to the tau-Higgs Yukawa coupling and defining $\chi_l =
(M_X/M_0)^{-h_{\tau}^2/16 \pi^2}$ we obtain:
\begin{equation}
{h_{\nu_{e,\mu}}\over h_{\nu_{\tau}}}\, (M_X) \simeq \chi_l\,
{h_{\nu_{e,\mu}}\over h_{\nu_{\tau}}} (M_0) \,,
\end{equation}
and:
\begin{equation}
{h_{e,\mu}\over h_{\tau}}\, (M_X) \simeq \chi_l^3 \,
{h_{e,\mu}\over h_{\tau}}\, (M_0) \,.
\end{equation}
Notice the difference in the evolution of the up-type and down-type lepton-mass
ratios to the one in the quark sector.
Again in the presence of a predominantly large third-generation Yukawa
coupling also the lepton-mixing angles undergo a slow evolution:
\begin{equation}
{d \over dx}\, ln\mid V_{13}\mid\, \simeq\,
{d \over dx}\, ln\mid V_{23}\mid\, \simeq \, h_{\tau}^2 \,,
\end{equation}
while the evolution of $\mid V_{12} \mid$ is negligible as in the quark sector.
Since the numerical value of $\chi_l$ is so close to one and given the
uncertainty stemming from the unknown Majorana sector, we conclude that
radiative corrections to the neutrino masses and mixings are less important in
this context and will be neglected.

\section{Implications for neutrino-oscillation experiments.}

Our previous discussion of the dependence of the light-neutrino mass matrix on
the structure of the heavy Majorana sector has revealed
at least four distinct classes of neutrino spectra, that span
the range of predictions expected from realistic GUT models, that satisfy the
principle of economicity.
The question now is how to distinguish between models that belong to different
classes, due to a different quark-Yukawa or/and heavy Majorana sector,  with
the help of neutrino-oscillation experiments.
For this, one first needs to know the kind of neutrino-flavour  transition that
is most likely to occur in each case in order to decide
on the type of experiments that are best suited.

In ref.(\cite{Petcov}), it was shown, that, if there is hierarchy among the
neutrino masses the largest transition probability will be into the
heaviest neutrino:
\begin{equation}\label{TP}
P (\nu_i\to \nu_j) \simeq \delta_{ij} - 2 \mid V_{\nu_3 - j}\mid^2
\, (\delta_{ij} - \mid V_{\nu_3 - i}\mid^2) \times
[1 - cos ({\Delta m^2 L\over 2 p})] \,,
\end{equation}
where $p$ and $L$ are the neutrino momentum and distance from the source to the
detector, and $\Delta m^2$ $\simeq m_{\nu_3}^2 - m_{\nu_1}^2$
$\simeq m_{\nu_3}^2 - m_{\nu_2}^2 \sim m_3^2$ is the only relevant mass
parameter. This implies that if the natural mass-hierarchy scenario of
equ.(\ref{NH}) is realised in nature, as predicted by GUT models classified
under our cases {\bf (i)} and {\bf (ii)}, the transition will be preferably
into the tau neutrino. On the other hand, if an inverse mass-hierarchy scenario
according to eqs.(\ref{RH1},\ref{RH2}) is present, as suggested
by models obeying the conditions that define case {\bf (iv)} and partly case
{\bf (iii)}, the transition will be predominantly into the muon neutrino.

Another consequence of equ.(\ref{TP}) is the existence of
hierarchy in the transition probabilities as a result
of the hierarchy of the mixing-matrix elements.
Therefore, for models belonging to the classes (I - IV) and fulfilling in
general the
conditions of case {\bf (i)} or {\bf (ii)} - except for the special case
where $\mid V_{\nu_e - \tau}\mid$ -, the hierarchy of the mixing elements,
eqs.(89,94), implies the following hierarchy of transition probabilities:
\begin{equation}
P^{(i);(ii)}_{(I - IV)} (\nu_e\to \nu_{\mu}) \ll
P^{(i);(ii)}_{(I - IV)} (\nu_e\to \nu_{\tau}) \ll
P^{(i);(ii)}_{(I - IV)} (\nu_{\mu}\to \nu_{\tau}) \,.
\end{equation}
So the best way to test these classes of models is to look for
$\nu_{\mu}\leftrightarrow \nu_{\tau}$ oscillations.
Given the present experimental sensitivity \cite{E56}, one can deduce
from the fact that no $\nu_{\mu}\to \nu_{\tau}$ oscillations have been observed
that the mass of the tau neutrino cannot lie above a few eV.
On the other hand,
for models belonging to class (V) the transition probabilities will always
satisfy the anomalous pattern:
\begin{equation}
P_{(V)} (\nu_e\to \nu_{\mu}) \ll
P_{(V)} (\nu_{\mu}\to \nu_{\tau}) \ll
P_{(V)} (\nu_e\to \nu_{\tau}) \,,
\end{equation}
and $\nu_e\leftrightarrow \nu_{\tau}$ oscillation
experiments would have been the best place to look for them. Unfortunately,
the sensitivity of present and planned experiments \cite{Los
Alamos} is off the range predicted by these models.

What are the experimental prospects for the future?
The range of $\Delta m^2$ and $sin^2 2\theta$ that could be explored for
$\nu_{\mu}\leftrightarrow \nu_{\tau}$ oscillations with the two CERN
experiments CHORUS
and NOMAD that are scheduled for next year is \cite{Rubia}:
\begin{equation}
sin^2 2\theta_{\nu_{\mu}\leftrightarrow \nu_{\tau}} \geq 2.3 \times 10^{-4}
\quad {\rm for} \quad \Delta m^2 \geq
(7 eV)^2 \,,
\end{equation}
and $\Delta m^2$ $\simeq 2\times 10^{-1} eV^2$ for maximal mixing.
With respect to GUT models belonging to the classes (I - IV) that predict
a natural or an only slightly distorted neutrino-mass hierarchy (our cases {\bf
(i)} and {\bf (ii)}), these experiments represent a very exciting testing
ground. If the tau-neutrino mass is of the order of a few electronvolts, then
$\nu_{\mu} \leftrightarrow \nu_{\tau}$ oscillations should be measured in
accordance with eqs.(79 - 87) and the values given in table (3) respectively.
Then the solar neutrino problem would be as well resolved, in terms of
matter-enhanced small-angle $\nu_e\leftrightarrow \nu_{\mu}$ oscillations.
This would indeed be the most satisfying scenario, solving simultaneously the
two neutrino-deficit problems.
It could however well be that the tau-neutrino mass is substantially below the
scale that is relevant to the solution of the dark matter problem, in which
case coming experiments will be insensitive to such oscillations and
the solar neutrino problem will not be resolvable, since the predicted mass
hierarchy is too large, eqs.(72,73,93).
This of course does not exclude such models. On the contrary, it leaves us
with the option that the scale of the heavy Majorana sector is closer to
$M_G$, {\it i.e.} $10^{14} - 10^{16}$, a possibility, which looks in fact more
natural for SUSY GUT models that do not contain an intermediate scale.
Even if the sensitivity to this transition could be increased by an order of
magnitude with the ICARUS detector, that is planned to settle the dispute on
the atmospheric-neutrino deficit and distinguish between the
three possible solutions to the solar-neutrino deficit \cite{Rubia},
this would not significantly improve the testing of this type of models.
On the other hand, there are currently also long-range oscillation experiments
with the CERN $\nu_{\mu}$ beam send to Gran Sasso and/or Superkamiokande under
investigation \cite{Rubia}, that could reach $\Delta m^2 \sim 10^{-4} eV^2$ for
full mixing in vacuum and push the Majorana scale beyond $M_G$, bringing GUT
models with a single Majorana scale into difficulties.

On the other hand, one can with the sensitivity reached by experiments like the
CDHS and CHARM, that have searched for $\nu_{\mu}\leftrightarrow \nu_{\tau}$
oscillations,
set a limit of $m_{\nu_{\mu},\nu_{\tau}} \leq 2. - 0.5$ eV for
$sin^2 2\theta_{\mu \tau} \simeq 0.1 - 1$ for models belonging to case {\bf
(iii)}, that predict a strong mixing between the muon and tau neutrino. The
corresponding limits for models belonging to case {\bf (iv)} that predict
maximal mixing between $\nu_{\mu} - \nu_e$, equ.(104), and between $\nu_e -
\nu_{\tau}$, equ.(105), are: $\Delta m^2 \leq 7.\times 10^{-2} eV^2$ and
$\Delta m^2 \leq 2.\times 10^{-2} eV^2$ respectively \cite{Acclim},
\cite{Zacek}. Of the cited limits the first two exclude
the possibility that any combination of the SM neutrinos can resolve the HDM
problem, while the third limit leaves us with the option that the muon neutrino
is the HDM candidate. Maximal mixing between $\nu_e - \nu_{\mu}$ or
between $\nu_e - \nu_{\tau}$ could also be the explanation for the observed
solar neutrino deficit according to the large-angle MSW or the vacuum
oscillation solution. It is interesting to note that the expected sensitivity
of future $\bar{\nu}_e \leftrightarrow \bar{\nu}_x$ experiments is getting
close to testing the first of these two options.
A detailed account of the
potential contained in cases {\bf (iii)} and {\bf (iv)}, as far as theory and
experiment is concerned, will be given elsewhere.

\section{Conclusions}
The new classification scheme of supersymmetric grand-unified models,
that has emerged from the requirement of having a
most economical quark-Yukawa sector at the unification scale \cite{RRR},
has been extended also to the lepton sector such as to include neutrino masses
and lepton mixing through the use of the
Georgi-Jarlskog relations and assuming the most general structure for the
heavy Majorana sector. The discussion of the latter has revealed yet another
classification scheme in terms of four distinct cases, that lead to universal
mass ratios and mixings for the three light neutrinos. The universality
manifests itself through the fact that models belonging to different classes
with respect to the structure of their quark- (and charged-lepton-) Yukawa
sectors can give the same neutrino spectrum if the heavy Majorana sector
satisfies certain requirements. The first case for example, which
is characterised by the presence of only one heavy Majorana scale and the
absence of any special symmetries for the heavy Majorana-mass matrix,
gives  neutrino-mass ratios that are typical for the quadratic seesaw, while
in the other cases, that are characterised by a hierarchy of heavy Majorana
scales and/or extra symmetries of the Majorana mass matrix, they get more
or less distorted up to the point that the natural mass hierarchy among the
generations can become reversed. In view of a possible testing of such
maximally-predictive GUT models through neutrino-oscillation experiments,
a comparision with existing and planned experiments has and could soon throw
some more light on the structure of the Yukawa interactions at energies
close to the grand-unification scale.

\vskip 1cm
\noindent
{\bf Acknowledgements:}
I would like to thank S.T. Petcov for interesting discussions. This work
was in part supported by the CEC Science Project $n^o$ SC1-CT91-0729.

\newpage
\pagestyle{empty}
\vskip 5cm
{\bf Table (1):} The values of the parameters in equ.(45) and equ.(46)
that correspond to the five distinct classes of maximally-predictive
GUT models from ref.[7].
\vskip 1cm
\begin{tabular}{|l|l|l|l|l|l|} \hline
& & & & & \\
& {\bf (I)} & {\bf (II)} & {\bf (III)} & {\bf (IV)} & {\bf (V)} \\
& & & & & \\
\hline
{\bf $\alpha$} & $\sqrt{2}$ & 1 & 0 & $\sqrt{2}$ & 0 \\
\hline
{\bf $\beta$} & 1 & 0 & 1 & $\sqrt{3}$ & $\sqrt{2}$ \\
\hline
{\bf $\gamma$} & 0 & 1 & 0 & 1 & 1/$\sqrt{2}$ \\
\hline
{\bf $\delta$} & 0 & 0 & $\sqrt{2}$ & 0 & 1 \\
\hline
$\alpha^{\prime}$ & 2 & 2 & 2 & 2 & 2 \\
\hline
$\beta^{\prime}$ & 2 & 2 & 2 & 2 & 2 \\
\hline
{\bf $\gamma^{\prime}$} & 4 & 2 & 4 & 0 & 0 \\
\hline
\end{tabular}

\newpage
\pagestyle{empty}
\vskip 5cm
{\bf Table (2) :}
The structure of the effective light-neutrino mass matrix to lowest order in z
for the five classes of models from ref.[7] in the case ${\rm M_R}$ fulfills
the conditions of equ.(57).
\vskip 10 mm
\begin{table}[h]
\centering
\begin{tabular} {|c|c|c|c|}
\hline
& & & \\
{Class} & {${\rm M}_{\nu}^{\rm eff} / {\rm m}_0$} & {${\rm
M}_{\nu}^{\rm eff} / {\rm m}_0$} & {Class} \\
& & & \\
\hline
& & & \\
{(I)} & $\pmatrix{2 a_1 z^6 & \sqrt{2} a_1 z^5 &
\sqrt{2} a_2 z^3 \cr \sqrt{2} a_1 z^5 & a_1 z^4 & a_2 {\rm
z}^2 \cr \sqrt{2} a_2 z^3 & a_2 z^2 & 1 \cr
}$ & $\pmatrix{\sqrt{2} z^4 & \sqrt{2} a_2 z^4 & \sqrt{2} z^2
\cr \sqrt{2} a_2 z^4 & a_1 z^4  & a_2 z^2 \cr
\sqrt{2} z^2 & a_2 z^2 & 1 \cr
}$ & {(III)} \\
& & & \\
\hline
& & & \\
{$\matrix { {\rm (II)} \cr {\rm + (IV)}\cr}$} & $\pmatrix{\alpha^2 a_1 z^6 &
\alpha a_2 z^4 & \alpha a_2 z^3 \cr \alpha a_2 z^4 & z^2  &
z \cr \alpha a_2 z^3 & z & 1 \cr}$ & $\pmatrix{z^4 & z^3 /
\sqrt{2} & z^2 \cr z^3 / \sqrt{2} & z^2 /2 & z/ \sqrt{2} \cr
z^2 & z / \sqrt{2} & 1 \cr
}$ & {(V)} \\
& & & \\
\hline
\multicolumn{4}{|c|}
{ } \\
\multicolumn{4}{|c|}
{$a_1 = {r_2 \over r_3} \  \ \ ; \ \ \ a_2 =
{r_6 \over r_3} \ \ \ ; \ \ \ a_3 = {r_4 \over r_3} \ \ \ ; \ \ \ a_4
= {r_5 \over r_3} \ \ \ ; \ \ \ a_5 = {r_1 \over r_3}$}\\
\multicolumn{4}{|c|}
{ } \\
\hline
\end{tabular}
\end{table}

\newpage
\pagestyle{empty}
\vskip 5cm
{\bf Table (3) :}
The lepton-mixing elements predicted by maximally-predictive GUT models,
belonging to classes (I - IV) from ref.[7], in the case {\bf (ii)} where
, due to the presence of symmetries in $M_R$, all but one of its
subdeterminants $r_1, r_2, r_4, r_5$ are zero.
\vskip 10 mm
\begin{table}[h]
\centering
\begin{tabular} {|c|c|c|c|c|}
\hline
& & & & \\ Mixing  & {\bf (I)} $ \begin{array}{cc}
& a_1 \not= 0 \\ {\rm or} & a_2 \not= 0 \\ {\rm or} &
a_3 \not= 0
\end{array} $ & $
 \begin{array}{rr} {\bf (I)} & \begin{array}{rr} & a_4 \not= 0 \\
                          {\rm or} & a_5 \not= 0 \end{array}  \\
                    {\bf (II)} & a_4 \not= 0 \\
                    {\bf (IV)} & a_4 \not= 0
\end{array} $ & $ \begin{array}{rr}
{\bf (II)}  & a_2 \not= 0 \\
{\bf (IV)} & a_2 \not= 0 \end{array} $ &
{\bf (III)} $ \begin{array}{cc} & a_1\not= 0  \\
{\rm or} & \alpha_3\not=0 \end{array} $ \\
& & & & \\
\hline
& & & & \\
$\mid V_{\nu_e-\mu}\mid$ & ${\lambda\over 3}$ & ${\lambda\over 3}$ &
$ 1 - {\lambda^2\over 18} $ & ${\lambda\over 3}$ \\
& & & & \\
\hline
& & & & \\
$\mid V_{\nu_{\mu}-\tau}\mid$ & $\gamma \lambda^3$ & $\gamma \lambda^3$ &
$0$ & $\gamma \lambda^3$ \\
& & & & \\
\hline
& & & & \\
$\mid V_{\nu_e-\tau}\mid$ & $\gamma \lambda^5$ & $0$ &
$ 2 \lambda^3$ & $\lambda^4$ \\
& & & & \\
\hline
\multicolumn{5}{|c|}
{ } \\
\multicolumn{5}{|c|}
{$ a_1 = {r_2 \over r_3} \  \ \ ; \ \ \ a_2 =
{r_6 \over r_3} \ \ \ ; \ \ \ a_3 = {r_4 \over r_3} \ \ \ ; \ \ \ a_4
= {r_5 \over r_3} \ \ \ ; \ \ \ a_5 = {r_1 \over r_3}$}\\
\multicolumn{5}{|c|}
{ } \\
\hline
\end{tabular}
\end{table}

\vskip 2cm

\begin{thebibliography}{11}
\bibitem{GG} M. Chanowitz, J. Ellis and M.K. Gaillard, \np 135,78,66.
\bibitem{GJ}H. Georgi and C. Jarlskog, \pl 86,79,297.
\bibitem{Harvey} J. Harvey, D. Reiss and P. Ramond, \pl 92,80,309;
\np 199,82,223.
\bibitem{DHR}S. Dimopoulos, L.J. Hall and S. Raby, \prl 68,92,1984; \prd
45,92,4192.
\bibitem{DHRN}S. Dimopoulos, L.J. Hall and S. Raby, \prd 47,93,3697;
L.J. Hall, {\bf UCB}-PTH-92-22.
\bibitem{Leonta}G.K. Leontaris and N.D. Tracas, \pl 303,93,50;
G.K. Leontaris and J.D. Vergados, \pl 305,93,242;
H. Dreiner, G.K. Leontaris and N.D. Tracas, {\bf OUTP}-92-27P.
\bibitem{RRR}P. Ramond, R.G. Roberts and G.G. Ross, {\bf RAL}-93-010.
\bibitem{Shafi} K.S. Babu and Q. Shafi, \pl 294,92,235.
\bibitem{Lang} V. Barger {\it et. al.},
\prd 47,93,1093;
P. Langacker and N. Polonsky, {\bf UPR}-0556T (1993);
M. Carena {\it et. al.}, {\bf MPI}-PH-93-10; W.A. Bardeen {\it et. al.}, {\bf
MPI}-PH-93-58; M. Carena {\it et. al.}, {\bf MPI}-PH-93-66.
\bibitem{discrete} B.R. Green {\it et. al.}, \np 278,86,667;
A. S. Joshipura and U. Sarkar, \prl 57,86,33;
G.G. Ross, \pl 200,88,441; T. Tamvakis, \pl 208,88,451; N. Ganoulis, G.
Lazarides and Q. Shafi, \np 323,89,374; N.I. Polyakov, \pl 255,91,77;
L.E. Ib\`{a}\~{n}ez and G.G. Ross, \pl 260,91,291; \np 368,92,3;
D.A. MacIntire, {\bf SCIPP}-93-27; M. De Montigny and M. Masip, {\bf
UFIFT}-93-11;
D.B. Kaplan and M. Schmaltz, {\bf UCSD}-PTH-93-30.
\bibitem{Giudice}G. F. Giudice, \journal Mod. Phys. Lett.,A7,92,2429.
\bibitem{Fritzsch}H. Fritzsch, \pl 70,77,436; \pl 73,78,317.
\bibitem{Stech} B. Stech, \pl 130,83,189.
\bibitem{RR}J. Ellis, S. Kelley and D.V. Nanopoulos, \pl 249,90,441;
P. Langacker and M. Luo, \prd 44,91,817;
U. Amaldi, W. de Boer and H. F\"urstenau, \pl 260,91,447;
R.G. Roberts and G.G. Ross, \np 377,92,571.
\bibitem{Elena} E. Papageorgiu and S. Ranfone, \pl 282,92,89.
\bibitem{Babu}K.S. Babu and R.N. Mohapatra, \prl 70,93,2845.
\bibitem{Ellis} J. Ellis, J.L. Lopez and D.V. Nanopoulos, \pl 292,92,189.
\bibitem{Langacker} S.A. Bludman, D.C. Kennedy and P.G. Langacker, \prd
45,92,1810;
M. Cvetic and P. Langacker, \prd 46,92,2759; P. Langacker, talk given at the
Unified Symmetry in the Small and in the Large, Coral Gables, Florida, January
1993.
\bibitem{Rubia} N. Armenise {\it et. al.}, {\bf CERN}-SPSC/90-40, SPSC P254 and
PPE int. report 1993; L. DiLella, {\it in} Proc. Neutrino 92, \np 31,93,319;
see also C. Rubia, {\bf CERN-PPE}/93-08 and references therin.
\bibitem{Spiro} For a recent review, see {\it e.g.} M. Spiro, talk given at the
High Energy Physics Conference, Marseille, July 1993.
\bibitem{seesaw} M. Gell-Man, P. Ramond and R. Slansky, {\it in} Supergravity,
ed. P. van Niewenhuizen and D.Z. Freedman (north Holland, Amsterdam, 1979);
T. Yanagida, {\it in} Proc. Workshop on the Baryon number of the universe and
unified theories, Tsukuba, Japan, 1979, ed. O. Sawada and A. Sugamoto (KEK,
Tsukuba, 1979).
\bibitem{USS}A. Davidson and K.C. Wali, \prl 58,87,2623;
{\bf 59}(1987)393; S. Ranfone, \prd 42,90,3819;
E. Papageorgiu and S. Ranfone, \np 369,92,99.
\bibitem{Petcov} S.M. Bilenky, M. Fabbrichesi and S.T. Petcov, \pl 276,92,223.
\bibitem{Ga} R. Davis {\it et. al.}, {\it in} Proc. of the 21st International
Cosmic Ray Conference, Vol. 12, ed. R.J. Protheroe (University of Adelide
Press, Adelide, 1990);
K.S. Hirata {\it et. al.}, \prl 65,90,1297;
A.I. Abazov {\it et. al.}, \prl 67,91,3332;
P. Anselman {\it et. al.}, \pl 285,92,376.
\bibitem{SSM} J.N. Bahcall and R.K. Ulrich, \rmp 60,88,297;
S. Turck-Chieze {\it et. al.}, \journal Astrophys. J.,335,88,415; J. H.
Bahcall and M. Pinsonneault, \rmp 64,92,885.
\bibitem{Bludman} S.A. Bludman, N. Hata, D.C. Kennedy and P. Langacker,
{\bf UPR}-0516 (1992); X. Shi, N.D. Schramm and J. Bahcall, \prl 69,92,717.
\bibitem{Ponte} B. Pontecorvo, \journal Zh. Eksp. Teor. Fiz.,33,57,549;
{\bf 34}(1958)247; {\bf 53}(1967)1717.
\bibitem{MSW} L. Wolfenstein, \prd 17,78,2369; {\bf D20} (1979) 2634; S.P.
Mikheyev and A.Y. Smirnov, \sovj 42,86,1441.
\bibitem{Krastev}P.I. Krastev and S.T. Petcov, \pl 299,93,99.
\bibitem{Smirnov}A.Y. Smirnov, D.N. Spergel and J.N. Bahcall, {\bf
IASSNS}-AST-93-15;
\bibitem{IMB} Ch. Berger {\it et. al.}, \pl 245,90,305;
M.M. Boliev {\it et. al.}, {\it in} Proc. of the 3rd Int. Workshop on
Neutrino Telescops, ed. M. Baldo-Ceolin, 1991;
K.S. Hirata {\it et. al.}, \pl 280,92,146;
R. Becker-Szendy {\it et. al.}, \prd 46,92,3720;
M. Goodman, {\it at} APS-DPF Meeting, Fermilab, November 1992.
\bibitem{Silk} See, {\it e.g.} talk of J. Silk at the High Energy Physics
Conference, Marseille, July 1993.
\bibitem{Acclim} L.A. Ahrens {\it et. al.}, \prd 25,85,2732;
B. Blumenfeld {\it et. al.}, JHUHEP 1289-1 (1991).
\bibitem{Mark} K. Enquist, K. Kainulainen and M. Thomson, \pl 280,92,245.
\bibitem{Caldwell}D.O. Caldwell and R.N. Mohapatra, {\bf UCSB}-HEP-93-03.
\bibitem{Danilov} M. Danilov, talk given at the High Energy Physics Conference,
Marseille, July 1993.
\bibitem{GSTO}R. Gatto, G. Sartori and M. Tonin, \pl 28,68,128;
R.J. Oakes, \pl 29,69,683; \pl 30,70,262.
\bibitem{Pokorski} M. Olechowski and S. Pokorski, \pl 257,91,388;
\pl 231,89,165.
\bibitem{Wolfenstein} L. Wolfenstein, \prl 51,83,1945;
P. Fishbane and P.Q. Hung, \prd 45,92,293;
Y. Koide, H. Fusaoka and C. Habe, \prd 46,92,4813.
\bibitem{oldGUT} See {\it e.g.}, R.N. Mohapatra and P.B. Palash, Massive
Neutrinos in Physics and Astrophysics, World Scientific Lecture Notes in
Physics, Vol. 41, 1991.
\bibitem{Antoniadis} I. Antoniadis and G.K. Leontaris, \pl 216,89,333;
I. Antoniadis, J. Ellis, J.S. Hagelin and D.V. Nanopoulos, \pl 194,87,231; \pl
205,88,450; \pl 208,88,209; \pl 231,89,65;
I. Antoniadis, G.K. Leontaris and J. Rizos, \pl 245,90,161;
G.K. Leontaris, J. Rizos and K. Tamvakis, \pl 243,90,220.
\bibitem{Ranfone} G.K. Leontaris and J.D. Vergados, \pl 258,91,111;
S. Ranfone, \pl 298,93,351;
S. Ranfone and E. Papageorgiu, \pl 295,92,79.
\bibitem{nonren} S. Nandi and U. Sarkar, \prl 56,86,564;
M. Cvetvic and P. Langacker, \prd 46,92,2759;
S. Kalara, J.L. Lopez and D.V. Nanopoulos, \pl 245,90,421; \np 353,91,650;
J.L. Lopez and D.V. Nanopoulos, \pl 251,90,73; A.E. Faraggi, \np 403,93,101.
\bibitem{E56} N. Ushida {\it et. al.}, \prl 57,86,2897.
\bibitem{Los Alamos} C. Angelini {\it et. al.}, \pl 179,86,307;
J. Schneps, {\it in} Neutrino 92, \np 31,93,307.
\bibitem{Zacek} G. Zacek {\it et. al.}, \prd 34,86,2621;
G.S. Vidyakin {\it et. al.}, \jetp 98,90,764.
\end{thebibliography}
\end{document}